\documentclass[11pt,showpacs,showkeys,nofootinbib,preprintnumbers]{revtex4-2}
\usepackage{graphicx}   
\usepackage{color}         
\usepackage{braket}
\usepackage{amsmath}
\usepackage{amssymb}
\usepackage{titlesec}
\usepackage{makecell}
\usepackage{natbib,hyperref}
\usepackage[mathscr]{euscript}
\usepackage{dutchcal}

\def \bea {\begin{eqnarray}}
\def \eea {\end{eqnarray}}

\newcommand{\de}{\partial}

\usepackage{mathrsfs}  
\usepackage{bm}

\definecolor{ao-english}{rgb}{0.0, 0.5, 0.0}
\definecolor{cadmiumblue}{rgb}{0.0, 0.42, 0.24}

\newcommand{\nn}{\nonumber}

\begin{document}

\preprint{ECTP-2025-07}
\preprint{WLCAPP-2025-07}
\hspace{0.05cm}

\title{The Derivation of Phase-Space Metric in a Geometric Quantization Approach: General Relativity with Quantized Phase-Space Metric and Relative Spacetime}
\email{Corresponding author: atawfik@acu.edu.eg; 400778@iu.edu.sa; atawfik@bnl.gov}

\author{Kareem~Mubaidin$^{1,2}$, 
Debarshi~Mukherjee$^{3}$, 
Saleh~O.~Allehabi$^{4}$, 
Azzah~A.~Alshehri$^{5}$, 
Mahmoud~Nasar$^{6}$, 
Ashraf F. El-Sherif$^{7}$, 
Abdel Nasser~Tawfik$^{4}$}

\affiliation{$^1$Egyptian Center for Theoretical Physics (ECTP@ACU), October Gardens, 12556 Giza, Egypt}
\affiliation{$^2$International School of Choeuifat Cairo, 11511 New Cairo, Egypt} 
\affiliation{$^3$Indian Institute of Science Education and Research (IISER) Berhampur, Vigyanpuri, Laudigam, Konisi, Berhampur, 760003, Odisha, India} 
\affiliation{$^4$Department of Physics, Faculty of Science, Islamic University of Madinah (IU), Madinah 42351, Saudi Arabia} 
\affiliation{$^5$Department of Science and Technology, University College at Nairiyah, University of Hafr Al Batin (UHB), Nairiyah 31981, Saudi Arabia} 
\affiliation{$^6$Physics Department, Faculty of Science, Benha University, 13511 Benha, Egypt} 
\affiliation{$^7$Basic Science Department, Faculty of Engineering, Ahram Canadian University (ACU), 12451 Giza, Egypt}

\begin{abstract}
Various extensions to Riemann geometry have been proposed since the inception of general relativity (GR). The aim has been and continues to be to construct a quantum and dynamic spacetime that incorporates the well-known classical (static) spacetime. Apparently, this seems to enable the principles of GR and quantum mechanics (QM) to be reconciled into a coherent relativity and quantum theory. A canonical geometric quantization approach that presents kinematics of free-falling quantum particles within a tangent bundle, expands QM to incorporate relativistic gravitational fields, and generalizes the four-dimensional Riemann manifold into an eight-dimensional one likely discretizes, if not fully quantizes, the Finsler and Hamilton structures. The Finsler and Hamilton metrics can be directly derived from the Hessian matrix. As introduced in [Physics, 7 (2025) 52], the quantized four-dimensional metric tensor can be deduced by means of approximations including proper parameterization of coordinates and the equating line elements on all these manifolds including Riemann manifold. This research, on the contrary, goes beyond all these approximations and proposes the incorporation of a phase-space metric tensor into  GR. The derivation of a quantized eight-dimensional metric tensor is not only presented, but also the implications of it and the corresponding relative spacetime are examined.
\end{abstract}


\maketitle

\tableofcontents

\section{Introduction}

The quest to reconcile the principles of general relativity (GR) and quantum mechanics (QM)  has been, and continues to be, one of the most significant challenges in fundamental science. \cite{RevModPhys.17.120,born1938suggestion,Overduin:1997sri,Donoghue:1996ma,Maccone:2018rxs}. Although Einstein's GR has proven successful in integrating a classical approximation of spacetime to describe a variety of phenomena, particularly at large scales, and QM has excelled in explaining numerous low-scale phenomena, both fundamental theories continue to be fundamentally incompatible \cite{TEITEL201941,ROMERO2019180}. Einstein's GR is based on the assumption of a smooth and certain spacetime, in contrast to QM, which incorporates roughness and uncertainty. This essential contradiction holds particular importance in black hole physics and the analysis of other extreme phenomenon, including cosmological \cite{Singh:2026dyn,Chokyi:2026ryf,Singh:2026xin,} and spatial singularities \cite{Nordstrom1914,Weyl:1918ib,Stewart:1990ngn,Mannheim:2011ds,Tawfik:2026pcn} as well as quantizing stress-energy tensor \cite{Tawfik:2026ufr}. A recent quantization approach was established on extending spacetime geometry into phase-space geometry, which allows for the integration of momentum space \cite{Tawfik:2025icy,Tawfik:2025rel,Tawfik:2025kae,Tawfik:2024itt,NasserTawfik:2024afw,Tawfik:2023ugm,Tawfik:2023hdi}. This new approach utilizes the Relativistic Generalized Uncertainty Principle (RGUP), a physically motivated generalization of QM that incorporates relativistic gravitational fields and introduces modifications to the momentum operator \cite{Todorinov:2020jtq,Tawfik:2024gxc,Tawfik:2023onh,Tawfik:2023orl}. 

As mentioned, the different representations of gravity and spacetime in both theories, GR and QM, are the main reasons for their orthogonality. As stated, while QM doesn't describe gravity (neither classical nor quantum), the spacetime of GR is yet to be quantized \cite{Kiefer2020}. In other words, while QM covers all fundamental forces with the exception of gravitation, which is essential in GR. The geometric quantization method suggests a means to harmonize the fundamental principles of both theories. It presents two mutually dependent generalizations. It introduces two interconnected generalizations. This includes {\it ''gravitizing QM''} \cite{Penrose:2014nha}, for instance, through the application of the RGUP, the four-dimensional relativistic approach \cite{Todorinov:2020jtq,Tawfik:2024gxc,Tawfik:2023onh,Tawfik:2023orl} or its three-dimensional classical counterpart, the generalized uncertainty principle (GUP) \cite{Tawfik:2014zca,Tawfik:2015rva}. Conversely, in GR, spacetime and geometry are considered fundamental degrees of freedom, and assumed to be either static or, at most, undergoing gradual alterations \cite{Snyder:1946qz}. It is apparent that spacetime would not continue to be smooth and certain, particularly at relativistic energies and in finite gravitational fields \cite{bronshtein1936quantization}. 

One of the authors (AT) suggested the quantization of GR by converting the fundamental geometry into a dynamic form. Consequently, the mathematical depiction of spacetime tends to be either background dependent, discretized, or ultimately quantized. In this approach, the metric tensor plays a crucial role; being essential for the Einstein-Hilbert action and remaining invariant under diffeomorphisms \cite{Tawfik:2025icy,Tawfik:2025rel,Tawfik:2025kae,Tawfik:2024itt,NasserTawfik:2024afw,Tawfik:2023ugm,Tawfik:2023hdi}. The primary notion here is centered on introducing invariant commutation relations to QM, stemming from finite gravitational fields in vacuum. Therefore, the motion of gravitational quanta acts as means by which gravitational interaction takes place  \cite{Tawfik:2025icy,Tawfik:2025rel,Tawfik:2025kae,Tawfik:2024itt,NasserTawfik:2024afw,Tawfik:2023ugm,Tawfik:2023hdi}. Given that the metric tensor combines all necessary information about geometry and spacetime
\cite{BeltranJimenez:2017uwv,Tawfik:2023hdi,Tawfik:2023ugm}, the quantization of GR is therefore based on the quantization of its metric tensor \cite{Tawfik:2025icy,Tawfik:2025rel,Tawfik:2025kae,Tawfik:2024itt,NasserTawfik:2024afw,Tawfik:2023ugm,Tawfik:2023hdi}. The metric to be derived within Finsler and Hamilton geometry, upon which quantum mechanical modifications are applied, can be converted into the conventional Riemann metric. This constraint is intended to maintain all the assumptions of Einstein's GR, including symmetry and dimensionality of the metric tensor.

Starting with the geometric quantization on a four-dimensional manifold, which arises from the tangent or cotangent bundle, it is appropriate to first examine the quantization of the metric tensor in such eight-dimensional bundles. Concretely, the aim is to derive the quantized metric tensor on the cotangent bundle $T^{*}_x M$ over spacetime with coordinates $X=[x^{\alpha}, p_{\beta}]$. It is clear that this metric depends not only on the spacetime coordinates $x^{\alpha}$, but also on the conjugate momenta $p_{\beta}$.

This manuscript is structured as follows. The four-dimensional quantized metric tensor is briefly reviewed in Section \ref{sec:4frml}. 
The relativistic generalized uncertainty principle (RGUP), the approach which introduces gravitation to QM, is reviewed in Section. \ref{sec:RGUP}. 
The generalization of GR via extending Riemann to Finsler geometry is discussed in Section  \ref{sec:RiemannGen}. A brief review of quantizing the Finsler structure is elaborated in Section \ref{sec:qFnslt}. The mathematical approach to extend Finsler to Hamilton geometry is summarized in Section \ref{sec:qHmln}.  The quantized metric tensor in four-dimensional Riemann geometry is discussed in Section \ref{sec:tgmunu2}. The derivation of quantized metric tensor in phase space geometry is introduced in \ref{sec:tgmunu3}.  Section \ref{sec:Imlcs} is devoted to implications of the quantized phase-space metric to various aspects of GR. This includes line element, geodesic equations, Ricci curvature tensor and scalar, and Einstein field equations in Section \ref{sec:lineElmnt}, \ref{sec:geodsc}, \ref{sec:CrvTnsr}, \ref{sec:EFE8D}, respectively. 
The final conclusions are outlined in Section \ref{sec:cncls}.

\section{Brief Review on Quantizing Metric Tensor in Four-Dimensional Riemann Geometry}
\label{sec:4frml}

Let us start by introducing a brief review of the relativistic generalized uncertainty principle (RGUP), Section \ref{sec:RGUP}. RGUP is assumed to extend QM to integrate relativistic gravitational fields \cite{Todorinov:2020jtq,Tawfik:2024gxc,Tawfik:2023onh,Tawfik:2023orl}. 

\subsection{QM Generalization: Relativistic Generalized Uncertainty Principle}
\label{sec:RGUP}

Through the integration of quantum gravity influences in Hilbert space
\cite{Berglund:2022skk,Freidel:2005me} and the implementation of a physical mechanism to control the collapse of the wave function
\cite{Penrose:2014vok,Diosi2013,DIOSI1987377}, QM is assumed to be generalized. As discussed briefly, a third approach is introduced by the RGUP \cite{Todorinov:2020jtq,Tawfik:2024gxc,Tawfik:2023onh,Tawfik:2023orl}. The Heisenberg uncertainty principle (HUP), an fundamental theory of QM, emphasizes the noncommutative relationships that exist, for instance, between position $\hat{x}$ and momentum operators $\hat{p}\gtrsim \hbar$ within flat spacetime, particularly in the presence of a negligible gravitational field. The incorporation of a finite gravitational field and four-dimensional spacetime into the HUP integrates gravity and leads to the RGUP \cite{Todorinov:2020jtq,Tawfik:2024gxc,Tawfik:2023onh,Tawfik:2023orl}. 

The reason why RGUP and not the Generalized Uncertainty Principle (GUP) is preferred is that GUP seems to be less significant when evaluating the Universe on a large scale. Further reasons can be retrieved from Ref. \cite{physics7040052}. The thought experiment, which Heisenberg himself devised within an expanding background geometry, can be regarded as the foundation for the generalization of the Heisenberg Uncertainty Principle (HUP), known as the Extended Uncertainty Principle (EUP) \cite{Park:2007az,Bambi:2007ty}
\bea
\Delta x_i \Delta p_j \geq \hbar \left[1+\beta^2_2 \frac{\left(\Delta x_i\right)^2}{\ell^2}\right],
\eea
where $\ell$ represents the correct spatial uncertainty \cite{Zhu:2008cg}, and $\beta_2$ is a dimensionless quantity that can be determined empirically \cite{Diab:2020jcl}. The scale factor can thus be utilized to locate the position $x$. Moreover, the Hubble horizon sets the upper limit for the EUP approach. Hence, the proper generalization of QM that can be applied to the geometric quantization approach would integrate both the GUP and the EUP. In this regard, the minimal length produced by both GUP and EUP is (i) frame dependent \cite{Snyder:1946qz}. (ii) does not entirely preserve Lorentz covariance \cite{Szabo:2006wx} and (iii) appears to distort the Einstein dispersion relation \cite{Todorinov:2020jtq}. This distortion contradicts the linear addition rule of momenta \cite{Amelino-Camelia:2014gga}.

Thus, we conclude that the appropriate HUP extension that suites well the cosmological aspects, such as curved spacetime, must be four-dimensional and relativistic. One must first acknowledge that the position and momentum coordinates associated with a free particle in curved spacetime are not considered canonically conjugate variables \cite{Todorinov:2020jtq}. We focus on the physical position and momentum coordinates of the free particle and investigate their noncommutative relationships in four-dimensional curved spacetime, where finite relativistic and gravitational fields are present. Since these two quantities can be expressed in terms of their auxiliary four-vectors, specifically $x_{0}^{\mu}$ and $p_{0}^{\nu}$, one finds that $x^{\mu}\left(x^{\mu}_0, p_0^{\mu}\right)$ and $p^{\mu}\left(x^{\mu}_0, p_0^{\mu}\right)$. The canonically conjugate quantities $x_{0}^{\mu}$ and $p_{0}^{\nu}$ under the fundamental tensor $g^{\mu \nu}$ satisfy the noncommutative relation $\left[\hat{x}_0^{\mu}, \hat{p}_0^{\nu}\right] = i \hbar g^{\mu \nu}$, where $\hat{p}_0^{\mu}=-i\hbar \partial/\partial x_{0 \mu}$ \cite{Brandt:1991hy,phdthesisXun,Todorinov:2020jtq}. These assumptions make it possible to distinguish between the temporal and spatial components $x^{\mu} = (x_0^0, x_0^{i})$ and $p^{\mu} = \left(p_0^0, p_0^{i}\left(1+\beta p_0^{\rho} p_0^{\rho}\right)\right)$, where $x^0_0$ and $p^0_0$ are the temporal position operator and the $0$-th momentum operator, respectively \cite{Snyder:1946qz,phdthesisXun}. The stress-energy tensor for quantum particles in curved spacetime may be used to estimate both time components, namely $x^0_0=c t$ and $p^0_0 E/c$, where $t$ is the time and $E$ is the energy of free particle. The relationship between the physical position and momentum coordinates with respect to $x_{0}^{\mu}$ and $p_{0}^{\nu}$ was analyzed in refs. \cite{phdthesisXun,Todorinov:2020jtq}. In this regard, we emphasize that the proposed RGUP \cite{Tawfik:2023rrm} no longer suffers from (i) Lorentz covariance violation, (ii) coordinate dependency of the resulting minimum measurable length, and (iii) violation of the linear additive law of momenta. Therefore, RGUP reads \cite{Benczik:2002tt,Todorinov:2020jtq,phdthesisXun}
\bea
\left[\hat{x}^{\mu},\, \hat{p}^{\nu}\right] &=& i \hbar \left[\left(1 + \beta_3 p^{\rho} p_{\rho}\right) g^{\mu \nu} + 2 \beta_4 \hat{p}^{\mu} \hat{p}^{\nu}\right], \label{eq:4dxp}
\eea 
where $\beta_3$ and $\beta_4$ are two additional RGUP parameters. The HUP is easily obtained at $\beta_3=\beta_4=0$ as $\left[x^{\mu}, p^{\nu}\right] = i \hbar g^{\mu \nu}$. The validity of the noncommutative relations in curved spacetime is guaranteed by the resulting pseudo-Riemann metric tensor, $g^{\mu \nu}$. It is obvious that RGUP includes $g^{\mu \nu}$ and thereby becomes applicable on curved spacetime. Also it is crucial to emphasize that In isotropic discretized curved spacetime, the space and momentum operators in Eq. (\ref{eq:4dxp}) can be represented respectively as \cite{Snyder:1946qz}, 
\bea
\hat{x}^{\mu} &=& \hat{x}_0^{\mu}, \label{eq:4dx2} \\
\hat{p}^{\mu} &=& \left(1+\beta_3 p_0^{\rho} p_{0 \rho}\right) \hat{p}_0^{\mu}. \label{eq:4dp2}
\eea
This modification implies that the fundamental coordinates exhibit commutativity \cite{Tawfik:2014zca}. 
Throughout this work, operators $\hat{x}^{\mu}$ and $\hat{p}_{\nu}$ are promoted to effective phase-space coordinates $(x^{\mu}, p_{\nu})$ by taking expectation values in semiclassical states. Henceforth, $x^{\mu}$ and $p_{\nu}$ represent classical phase-space
variables unless stated otherwise.

Section \ref{sec:RiemannGen} reviews another generalization. Riemann geometry in GR is extended to Finsler geometry, which allows for the inclusion of quantum mechanical revisions.

\subsection{GR Generalization: Extending Riemann to Finsler Geometry}
\label{sec:RiemannGen}

The concept of a manifold and its underlying structure can be traced back to 1854, when Riemann expressed in his thesis the mathematical challenges that required resolution. An alternative resolution was proposed by Riemann himself suggesting that a smoothness approximation could be utilized. The resulting Riemann metric integrates the quadratic differential form along with a smoothly varying set of inner products in the tangent space \cite{bao2012introduction}. The quadratic differential within the context of Riemann geometry resulted in the concepts of sectional curvature and the curvature tensor. The general formulation of a metric was introduced by Finsler as the fourth root of a quartic differential form \cite{Finsler1918}. Characterized by a smoothly varying set of Minkowski structures, Finsler geometry is characterized by Finsler structure. Thus, the length of a curve segment $s$ between points $a$ and $b$ can be expressed as 
\bea
{\cal l}|_{\mathtt{Finsler}}=\int_a^b F\left(x^1,\cdots,x^n, y^1\cdots,y^n\right) dt,\qquad y=dx/dt,
\eea
where $x$ stands for position and $y$ for velocity. In this regard, we define the tangent space, at the point $x \in M$, as $T_x M$, with $M$ is an $n$-dimensional $C^{\infty}$ manifold. Therefore, the tangent bundle of $M$ is represented as $TM := \mathtt{U}_{x \in M} T_x M$. Every element of $TM$ is characterized by the structure $(x,y)$, where $x$ is an element of $M$ and $y$ is an element of $T_x M$. The natural projection $\pi:TM \rightarrow M$ is then expressed as $\pi(x,y) := x$. The cotangent bundle of $M$ is expressed as $T^{\ast} M := \mathtt{U}_{x \in M} T_x^{\ast} M$, where, at $x$, a cotangent space $T_x^{\ast}M$ of $M$ can be formed as the dual space corresponding to $T_x M$.

It has been indicated long time ago, that the essential geometry of spacetime can be expressed in terms of Finsler geometry \cite{Beem-1970}. However, its application should be limited by the cosmological principle confined to Finsler spacetime geometry that exhibits spatial homogeneity and isotropy \cite{Hohmann:2020mgs}. Assuming a proper parameterization, a measure for a curve on $M$ can be expressed as
\bea
s[\gamma] &=& \int_a^b F(\gamma(\tau), \dot{\gamma}(\tau)) d\tau, 
\eea
where $\gamma:\tau \mapsto \gamma(\tau)$ represents mapping of $[a, b]$ to $M$, whereas $\gamma:[a,b]\rightarrow M$ \cite{Hohmann:2018rpp}. As previously mentioned, Finsler geometry is characterized by: (i) a differentiable manifold $M$, and (ii) a continuous, non-negative structure $F:TM \rightarrow [0, +\infty)$ on the tangent bundle $TM$. The mapping $\pi:TM \rightarrow M$ indicates that for every point $X \in TM$, there exists a tangent vector to $M$, at the point $x \in M$, thereby establishing the relationship $X \mapsto x$. In other words, the Finsler spacetime is characterized as an eight-dimensional tangent bundle $TM$, which integrates the tangent spaces of the fundamental four-dimensional manifold $M$ \cite{Pfeifer:2011xi}.

The generalization of Riemann geometry is thus considered to generalize GR by relaxing the quadratic constraint on the line element $ds^2$. There are additional approaches that rely on such generalization, which include quantum gravity phenomenology \cite{Tavakol:2009zz}, relativistic gravity, and geometric axiomatic theories of spacetime and relativity \cite{Caianiello:1980iv,Brandt:1991sw,Tawfik:2023kxq}, and the violation of Lorentz invariance 
\cite{Girelli:2006fw}.

As discussed, the main purpose of generalizing Riemann geometry is the intention to impose quantum mechanical ingredients. Section \ref{sec:qFnslt} introduces a approach which allows for quantizing Finsler structure.

\subsection{Quantizing the Finsler structure}
\label{sec:qFnslt}

A real, nonnegative scalar function that establishes a canonical inner product but may not be generated by an inner product is known as the Finsler structure of an $n$-dimensional differentiable manifold $M$. The Finsler structure, which functions as a Minkowski norm, is implemented in every tangent space within $M$. The main distinction between Riemann and Finsler manifolds is that the Finslerian inner products are characterized by directions in the tangent bundle $TM$ rather than by points of $M$. This shows that the location and direction both have an impact on the inner product in a Finsler manifold. Furthermore, the quadratic restriction is relaxed in Finsler geometry, while this constraint is particular to Riemann geometry, in which the squared length, represented by the line element $ds^2$, is inherently quadratic with respect to the associated coordinate displacements. In addition to these differences, the Riemann and Finsler geometries share several similarities, including length, geodesics, curvature, connections, covariant derivative, and structural equations \cite{Yano1963,ASANOV2006275,Tawfik:2023orl}. The Finsler structure is expressed in terms of $2n$ independent variables, $F(x,y)$, with $x$ indicating the local coordinates and $y$ denoting the components of the contravariant vectors that are tangent to the manifold $M$, at the coordinates $x$. In relation to every tangent space $M_x$ connected to $M$, a conical domain $M^{\ast}_x$ can be recognized, wherein any tangent vector $y$ at the position $x$ is contained within $M^{\ast}_x$. Furthermore, all other tangent vectors that are collinear with $y$ and tangent at $x$ are similarly covered in $M^{\ast}_x$ \cite{bao2012introduction}. Due to its Finsler structure, the Riemann manifold is assumed to fulfill the following relation:
\bea 
F^2(x, y) &=& g_{\mu \nu}(x) y^{\mu} y^{\nu},  \label{eq:FnslrFRmn1}
\eea 
where $y=\dot{x}=dx/dt$. The Finslerian metric, $g_{\mu \nu}(x)$, is noticeably distinct from the Riemann metric $g_{\mu \nu}$. Equation \eqref{eq:FnslrFRmn1} distinctly illustrates the quadratic constraint placed on the line element within GR.

A pseudo-Riemannian manifold, an approach to quantize the line element, the fundamental first-form of curved spacetime, in the relativistic regime, was first presented in refs. \cite{Caianiello:1980iv,Caianiello:1989wm}, assuming that the additional curvatures associated with the eight-dimensional spacetime tangent bundle $TM=M_4\otimes M_4$ would be utilized to mimic the proposed quantization on the four-dimensional spacetime $M_4$. The quantization approach proposes extending the Riemann manifold of the GR to a Finsler manifold, which is a smooth $n$-dimensional differentiable manifold $M_4$ with a continuous nonnegative Finsler norm $F: TM \rightarrow [0, +\infty)$ on the tangent bundle. 

With the coordinates $x^{\mu}=(ct, x^i)$, where $\mu, \nu=0,1,2,3$ and $\dot{x}^{\nu}=\partial x^{\nu}/\partial s$ are tangent vectors and $\dot{x}^{\mu}\in T_xM$, where $T_{x}M$ represents the tangent bundle, at the coordinate $x$. Additionally, $TM:=U_{x\in M} T_xM$ indicates the tangent bundle across $M$. It is assumed that for every point $x$ on $M_{4}$, the mapping function $(x^{\mu}, \dot{x}^{\nu})\mapsto F(x^{\mu}, \dot{x}^{\nu})$ possesses the following properties:
\begin{itemize}
\item Positive definiteness, which means that the metric tensor is positive definite on $TM \backslash \{0\}$.
\item Positive homogeneity, which means that the relativistic four-velocity $\dot{x}^{\mu}$ makes $F$ positively $1$-homogeneous; $F(x^{\mu}, \lambda \dot{x}^{\nu})=\lambda\, F(x^{\mu}, \dot{x}^{\nu}),\;\; \forall \lambda \in R^+$, and
\item Triangle inequality, which is satisfied pointwise by $F(x^{\mu}, \vec{v}+\vec{w}) \leq F(x^{\mu}, \vec{v}) + F(x^{\mu}, \vec{w})$, where $\vec{v}$ and $\vec{w}$ are vectors tangent to $M_4$ at the point $x$, assuring subadditivity.
\end{itemize}

It is proper now to emphasize the features of Finsler manifold \cite{Tawfik:2015rva,Tawfik:2014zca,Tawfik:2016uhs,Diab:2020jcl}. These include:
\begin{itemize}
\item In contrast to the Finslerian manifold, which is defined by $M_4$ and $F(x^{\mu}, \dot{x}^{\nu})$, the Riemannian manifold is characterized by $M_4$ and $g_{\mu \nu}$. In this regard, the Finslerian manifold is characterized by: (i) a base space and (ii) a real scalar-valued function $F$. The base space is composed of set of points in $R^4$. The genuine scalar-valued function $R^4 \times R^4 \rightarrow R^+$ captures additional geometric structures.
\item Second, the components of the curvature tensor represent the non-commutation relations, whereas the conventional operators of the Heisenberg algebra are represented by the covariant derivatives on $TM$ \cite{Scarpetta2006,caianiello1979hermitian,caianiello1981extended}. 
\item Third, Finsler geometry is then characterized by quantifying distances in abstract spaces. In the context of the present approach to advance relativity theory and quantum geometry, the distance separating two events on a Finsler manifold is established in a way similar to the conventional Euclidean distance, defined as the length of the shortest path connecting them. The sum of the lengths of its infinitesimal line components, represented by $ds$, represents the length of a curve on a Euclidean manifold. In a Finslerian manifold, the sum of $ds$ is then weighted according to the position and direction. 
\end{itemize}
Hence, it is apparent that the directions not the magnitudes of the tangent vectors $\dot{x}^{\mu}$ define the Finsler geometry. 

Let us now introduce a modified momentum operator as follows:
\begin{equation}
    \tilde{p}_{\mu}=\phi(p)p_\mu,
\end{equation}
where the function $\phi(p)$ given as
\bea
\phi(p)=1+\beta p^\rho p_\rho,
\eea
is dimensionless so that $\beta$ has an inverse momentum squared dimensionality. As introduced in Section \ref{sec:RiemannGen}, the function $\phi(p)$ is derived in the RGUP, which generalizes QM to finite relativistic gravitational fields and thereby introduces roughness and uncertainty into the cotangent bundle. 

Furthermore, let us assume a quantum particle moving in the curvatures, that emerge due to the proposed quantization. Accordingly, the manifold $M$ shall be extended as follows:
\begin{equation}
\mathbf{F^2(x,p)}=F(x^\alpha, \tilde{p}^\beta), \label{eq:F2ext}
\end{equation}
where $\alpha, \beta \in \{0,1,2,3\}$. It is obvious that $F(x^\alpha, \tilde{p}^\beta)$ is homogeneous of degree one. Equation \ref{eq:F2ext} can be rather identified with Hamilton structure.

The Finsler metric,$\tilde{g}^{(F)}_{\mu \nu}$, is defined to assure non-quadratic and more general dependencies on $\dot{x}^\mu$. This metric can be derived from the Hessian matrix, 
\begin{equation}
    \tilde{g}^{(F)}_{\mu \nu}=\frac{1}{2}\frac{\partial^2 F^2}{\partial y^\mu \partial y^\nu}. \label{eq:Fhmlt1}
\end{equation}
This makes the metric positive definite on $TM/[0]$. Unlike the metric in Riemann geometry, the Finsler metric depends on both position and direction. 

Now let us start with the line element in Riemann geometry
\begin{equation}
    ds^2 = g_{ij} dx^i dx^j,
\end{equation}
which has a quadratic dependency on the position. The Finsler metric has non-quadratic dependency and is defined as
\begin{equation}
    F(x,y) = \sqrt{g_{ij}(x,y) y^i y^j},
\end{equation}
where $y^i=\dot{x}^i$ and $y^j=\dot{x}^j$. The utilization of Finsler geometry allows for dimensional and tensorial consistency as well as reparametrization properties. The functional length of a curve $L[s]$ is given by
\begin{equation}
    L[s]=\int^L_0 F(s(t),\dot{s}(t)) dt,
\end{equation}
which is invariant under parameterizations $t \rightarrow \tilde{t}$ when $\tilde{t}$ and $\dot{s}^\mu$ are inversely scaled. Using the positive homogenity of degree one, one may conclude that the Finsler function and momentum are linearly scaled
\begin{equation}
    m^{+}F(x, \dot{x}) = F(x.m^{+}\dot{x}),
\end{equation}
where $m^{+}$ is the positive mass and reparameterized quantity introduced to derive momentum.

As discussed earlier, the introduction of quantum particles in the additional curvatures is rather intended to extend Finsler to Hamilton geometry, Section \ref{sec:qHmln}.

\subsection{Extending Finsler to Hamilton Geometry}
\label{sec:qHmln}

Recall that the Finsler structure $F(x,\dot{x})$ scaled linearly with four-momentum $p_\mu=mg^{F}_{\mu \nu}\dot{x}^\nu$ which leads to the related Hamilton structure
\begin{equation}
    H(x,p)=\frac{1}{2}g^{\alpha \beta}_{(F)}(x,p)p_\alpha p_\beta. \label{eq:Hmltn}
\end{equation}
The Legendre transformation of the cotangent bundle metric $g^{\alpha \beta}_{(F)}(x,p)$ is considered under the condition that $L(x, \dot{x})=F(x,\dot{x})$.

Both $\dot{x}^\mu$ and $\dot{p_\mu}$ can be derived from the Hamilton equations, Eq. \eqref{eq:Hmltn},
\bea
    \dot{x}^\mu = \frac{\partial H}{\partial p_\mu}, \qquad && \qquad
    \dot{p_\mu} = -\frac{\partial H}{\partial x^\mu}. \label{eq:dxdp1}
\eea
Now, we discuss the mapping between the tangent bundle $(TM, \pi, M)$, Finsler, and the cotangent bundle $(T^{*}M, \pi^*, M)$, Hamilton. It is obvious that the generalized Finsler structure, Section \ref{sec:qFnslt}, is nonlinearly related to the generalized Hamilton structure, Eq. \eqref{eq:Hmltn}. As expressed in Eq. \eqref{eq:dxdp1}, momentum variables $p_{\mu}$ can be derived from $F^2$,
\bea
    p_\mu &=& \frac{\partial F^2}{\partial \dot{x}_\mu}. \label{eq:LgdrTr}
\eea
Then, the Legendre transform, Eq. \eqref{eq:LgdrTr}, leads to
\bea
    H(x,p) &=& p_\mu \dot{x}^\mu -F(x,\dot{x}).
\eea
Thus, we conclude that the Hessian of the Finsler function is non-singular on the slit tangent bundle.

By generalizing both GR and QM and introducing the kinematics of quantum particles in the additional curvatures, the quantized metric tensor can now be derived, Section \ref{sec:tgmunu2}.

\subsection{Quantized Metric Tensor in Four-Dimensional Riemann Geometry}
\label{sec:tgmunu2}

Assuming a suitable parameterization of the coordinates $x^{\mu} = x^{\mu}( \zeta)$ and momenta $p^{\mu} = p^{\mu}(\zeta)$ in both Finsler and Hamilton spaces. That is for the parameterizations $\zeta^{\mu}$, both sets of coordinates in the Finsler tangent (or Hamilton cotangent) bundle and on the Riemann manifold are linked. The metric tensor at that position, where both Riemann and Finsler/Hamilton manifolds are parameterized simultaneously, can be derived from the Hessian matrix of $F^2(x_0^{\mu}, \phi(\nu) p_0^{b})$,
\bea
g_{\mu \nu}(x) = \frac{1}{2} \frac{\partial^2}{\partial p_0^{\mu} \partial p_0^{\nu}} \phi^2 F^2(x_0^{\mu}, p_0^{\nu}). \label{eg:4Hessian1}
\eea 
It is obvious that the resulting $g_{\mu \nu}(x)$ is both positive and symmetric. In this context, it is important to remember that the coordinates on $TM$ are defined as $x^{\mu} = \left(x^{\mu}, p^{\mu}\right)=\left(x_0^{\mu}, \phi(p) p_0^{\mu}\right)$. The relationship between the metric and the Finsler/Hamilton structure is expressed as follows:
\bea
\phi^2(p) F^2(x_0^{\mu}, p_0^{\nu}) &=& g_{\mu \nu}(x) \dot{p}_0^{\mu} \dot{p}_0^{\nu}. \label{eq:4F2gab}
\eea 
It is apparent that Euler's theorem can be applied to Eq. (\ref{eg:4Hessian1}), so that $F^2(x_0^{\mu}, p_0^{\nu})$ leads to Eq. (\ref{eq:4F2gab}). From Eq. (\ref{eq:4F2gab}), it is found that, even if the constraint of a quadratic line element is not diminished, Finsler geometry still bears similarities to Riemann geometry. Thus, the metric tensor $g_{\mu \nu}(x)$ on the Finsler/Hamilton manifold seems to be distinguished from the conventional metric tensor $g_{\mu \nu}$ on Riemann manifold. The proper parameterization in Finsler/Hamilton space facilitates the identification of both metrics. To this end, let us utilize the line element on both $M$ and $TM$ manifolds so that
\bea
g_{a b}\, dx^{\mu}(\zeta)\, dx^{\nu}(\zeta) = \tilde{g}_{\mu \nu}\, d\zeta^{\mu}\, d\zeta^{\nu}. \label{eq:4L2OldCoord2}
\eea 

By limiting the discussion to the simplest metric on Finsler manifold, namely the Klein metric \cite{Mo2006}, which characterizes hyperbolic or projective geometry, where the geodesic equations are represented by Euclidean line segments \cite{Klein:1926tv}, the structure is defined as 
\bea
F(x_0, p_0) &=& \left[\frac{|p_0|^2 - |x_0|^2 |p_0|^2 + \langle x_0 \cdot p_0\rangle^2}{1-|x_0|^2} \right]^{1/2}. \label{eq:FKlein}
\eea
It is obvious that $|\cdot|$ and $\langle \cdot\rangle$ represent the standard Euclidean norm and inner product in $R^n$, respectively.  

In this context, it is important to remember that $F(x_0, p_0)$ is merely required to exhibit positive homogeneity of degree one, and there is no singular choice of Finsler structure. Various forms of $F$, including Randers, Kropina, Klein, and others, can all fulfill the homogeneity requirement.

When RGUP corrections are introduced, the generalization to the uncertainty principle leads to a deformation of spacetime at quantum scales, resulting in a metric that is contingent upon the selected Finsler structure. For instance, a Randers-type Finsler structure, which consists of both linear and quadratic terms, generates a metric characterized by anisotropic corrections, while a Kropina-type structure produces distinct singular behaviors. Consequently, the RGUP-corrected metric is not singularly defined; it is indicative of the specific ansatz for $F(x_0, p_0)$. Despite this, all RGUP-corrected metrics adhere to the homogeneity condition. Let us now summarize this discussion as follows: \begin{itemize} \item No uniqueness: RGUP corrections are structure-dependent.
\item The Klein metric, which is part of the Beltrami--Klein model of hyperbolic geometry, is characterized by being projectively flat and exhibiting constant negative curvature. It serves as a canonical example where modifications maintain projective flatness while deforming curvature.
\end{itemize}
By examining the differences between Klein and Randers corrections, one can demonstrate how various Finsler selections result in distinct quantum-modified metrics. The RGUP corrections seemingly modify the numerator and denominator in the Klein metric, Eq. \eqref{eq:FKlein}, in a manner that differs from that observed in a Randers-type structure.

As discussed, the translation of $F(x^{\mu}, \dot{x}^{\nu})$ into Hamilton structure is nonlinear. This process is founded on the Legendre transform. An alternative derivation is based on the homogeneity properties of $F(x^{\mu}, \dot{x}^{\nu})$. As introduced, RGUP modifies the conjugate momentum operator, allowing for the expression of $F(x^{\mu}, \dot{x}^{\nu})$ to be given as $F(x_0^{a}, \phi(p) p_0^{b})$. The resulting $F(x_0^{a}, \phi(p) p_0^{b})$ subsequently exhibits homogeneity of degree one in $p_0$.  This presumes that the function $\phi(p)=1+\beta p_0^{\rho}p_{0 \rho}$ is obviously homogeneous of degree zero in $p_0$. In this respect, let us recall that the RGUP parameter $\beta = \beta_0 G/(c^3 \hbar) = \beta_0(\ell_{\mathtt{p}}/\hbar)^2$, where $c$ is the speed of light, $\hbar$ is the Planck constant, $G$ is the gravitational constant, and $\ell_{\mathtt{p}}$ is the Planck length \cite{Tawfik:2015rva,Tawfik:2014zca,Tawfik:2016uhs,Diab:2020jcl}, and this does not depend on $x_0$. Consequently, $\beta$ exhibits $-2$-homogeneity and the function $\phi(p)$ is indeed $0$-homogeneous in $p_0$. This not only guarantees homogeneity, but also preserves other mathematical properties of the Finsler structure, $F(x^{\mu}, \dot{x}^{\nu})$, including smoothness, subadditivity, positive definiteness, and strong convexity, within the quantized Hamilton structure $F(x_0^{\mu}, \phi(p) p_0^{\nu})$.

The quantized fundamental metric tensor can now be obtained from Eqs. (\ref{eg:4Hessian1}) and (\ref{eq:4L2OldCoord2}) \cite{Tawfik:2023hdi,Tawfik:2023ugm,Tawfik:2023rrm,Tawfik:2023kxq}. The derivation details of the fundamental metric in Riemann geometry can be found in the Appendix \ref{sec:MtrRmn}. Presuming that $F(x_0^{\mu},  p_0^{\nu})$ can be associated with $F(x_0^{\mu}, \phi(p) p_0^{\nu})$, Eq. \eqref{eq:FKlein}, one finds that
\bea
\tilde{g}_{\mu \nu}  &=& \left(\phi^2(p)+2\frac{\kappa}{(p_0^0)^2} F^2\right) 
 \left[1 + \frac{\dot{p}_0^{\mu} \dot{p}_0^{\nu}}{{\mathscr F}^2} \left(1+2\beta p_0 ^{\rho} p_{0 \rho}\right) \right] g_{\mu\nu} \nn \\
&+& \left[\frac{d x_0^{\mu}}{d \zeta^{\mu}} \frac{d x_0^{\nu}}{d \zeta^{\nu}} + \left(1+2\beta p_0 ^{\rho} p_{0 \rho}\right) \frac{d p_0^{\mu}}{d \zeta^{\mu}} \frac{d p_0^{\nu}}{d \zeta^{\nu}} \right] d_{\mu\nu}(x), \label{eq:gmunuQ2}
\eea  
where $\kappa=\beta/(p_0^0)^2$. It is evident that $\dot{p}_0^{\mu} \dot{p}_0^{\nu}$ represents a squared proper force which is normalized against the maximum proper gravitational force, ${\mathscr F}$. This is a newly identified physical constant that emerges from the proposed geometrical quantization approach. It is clear that ${\mathscr F}$ is proportional to the maximum proper acceleration, ${\mathscr A}=(c^7/\hslash G)^{1/2}$ as discovered by Caianiello \cite{Caianiello:1981jq,caianiello1984maximal,Brandt:1988sh}. 

We would like to highlight that the second line of Eq. \eqref{eq:gmunuQ2} is presented as
\bea
d_{\mu\nu}(x) &=& 2 \frac{\kappa F^2}{(p_0^0)^2} \left\{ 4 \phi(p) \ell_{\mu} \ell^{\sigma} g_{\sigma \mu} - 4 \phi(p) F \ell^{\sigma} \left[\delta_{0 \nu} g_{\sigma\mu} + \delta_{0 \mu} g_{\sigma\nu}\right] 
+ K^2 \left[2+\phi(p)\right] \delta_{0 \mu} \delta_0^{\sigma} g_{\sigma\nu}\right\}. 
\eea  
To verify that the Riemann and Finsler measures of the line elements are the same, it is necessary to reformulate $d_{\mu \nu}(x)$ in terms of $g_{\mu \nu}$ as demonstrated in the first line of Eq. \eqref{eq:gmunuQ2}. This process still represents a substantial mathematical challenge that has not been resolved. Consequently, one is obliged to make a bold assumption by ignoring the entire second line in Eq. \eqref{eq:gmunuQ2}. There is a significant physical argument against the approximation that $d_{\mu\nu}(x)=0$. Let us begin with the complete representation of the symmetric tensor
 $d^{\mu \nu}$,
\bea
d^{\mu \nu}(x) &=& \left[ {\frac{4\, \kappa\, F^2\, \left(F^2\, \kappa + 2\, (p_0^0)^2\, \phi(p_0) \right) \left(2\, F^4\, g^{00}\, \kappa - 2\, F^2\, \kappa\, (p_0^0)^2 - (p_0^0)^4\, \phi(p_0) \right) }{(p_0^0)^2\, \phi^3(p_0)\, \left(6\, F^4\, g^{00}\, \kappa - 4\, F^2\, \kappa \, (p_0^0)^2 + (p_0^0)^4\, \phi(p_0)  \right) 
 \left(2\, F^2\, \kappa + (p_0^0)^2\, \phi(p_0) \right) }} \right]\times \ell^{\mu}\, \ell^{\nu} \nn \\
&+& \left[\frac{4\, \kappa\, p_0^0\, F^3\, \left(F^2\, \kappa + 2\, (p_0^0)^2\, \phi(p_0) \right)}{\phi^2(p_0)\, \left(6\, F^4\, g^{00}\, \kappa - 4\, F^2\, \kappa\, (p_0^0)^2 + (p_0^0)^{4}\, \phi(p_0) \right) \left(2\, F^2\, \kappa + (p_0^0)^2\, \phi(p_0) \right) } \right] \times \left[\ell^{\mu}\, g^{0\nu} + \ell^{\nu}\, g^{0\mu}\right] \nn \\
&+& \left[\frac{- 6\, F^4\, \kappa\, (p_0^0)^2}{\phi(p_0)\, \left( 6\, F^4\, g^{00}\, \kappa - 4\, F^2 \kappa\, (p_0^0)^2 + (p_0^0)^4\, \phi(p_0) \right) 
 \left(2\, F^2\, \kappa + (p_0^0)^2\, \phi(p_0) \right)}\right]\times g^{0\mu}\, g^{0\nu}. \label{eq:fab} \nn
\eea 
The assumption that $d_{\mu\nu}(x)=0$ concerns the function $\phi(p_0)$, which is not contingent upon either $x_0$. To summarize the correlation between vanishing $d{\mu \nu}(x)$ and vanishing $\phi(p_0)$:
\begin{itemize}
\item The quantization framework incorporates a correction term $d_{\mu \nu}(x)$ into the generalized metric tensor.
\item This correction is explicitly dependent on the function $\phi(p_0)$, which is associated with the auxiliary four-momentum and is essential for maintaining the special curvature characteristics of the Randers metric.
\item Specifically, the Randers metric necessitates that the second term (which includes $\phi(p_0)$) behaves as a 1-form that is linear in momentum. Should $\phi(p_0)$ be equal to zero, the correction term $d_{\mu \nu}(x)$ will also be zero, resulting in the retention of only the Riemannian component of the metric.
\end{itemize}
Consequently, vanishing $d_{\mu \nu}(x)$ results in  vanishing $\phi(p_0)$, as $d_{\mu \nu}(x)$ is directly constructed from $\phi(p_0)$. In the absence of $\phi(p_0)$, the quantized extension reverts to the classical Riemann metric, causing the unique Randers curvature characteristics to be forfeited. 

This manuscript is primarily focused on an approach to eight-dimensional metric, where the problematic of finite $d_{\mu \nu}(x)$ is not significant. The assumption $d_{
\mu \nu}(x)=0$, which is vital for deriving the four-dimensional metric, might be justified by limiting the analysis to the leading-order isotropic RGUP correction represented in the first term of Eq.  \eqref{eq:gmunuQ2}. In this context, the tensor $d_{\mu \nu}(x)$ can be interpreted as indicating additional anisotropic corrections that go beyond the primary conformal deformation. Consequently, the resulting metric would reflect a first-order truncation of the entire phase-space geometry, preserving the leading isotropic contribution while disregarding higher-order anisotropic effects. This explanation may offer a more coherent physical rationale for the approximation utilized in the manuscript. As previously stated, this aspect is not relevant to the current analysis.

However, the truncation of Eq. \eqref{eq:gmunuQ2} with the assumption that $d_{\mu\nu}(x)=0$ remains the only plausible approximation that evidently violates the finiteness of $\phi(p)$, particularly in the second line of Eq. \eqref{eq:gmunuQ2}. Given that the proposed quantization is still approximate, especially by $d_{\mu\nu}(x)=0$ and equating line elements on phase-space as well as on four-dimensional manifolds. However, the approximated quantization still enables us to present essential implications of the proposed geometric quantization. If the mathematical challenges in re-expressing $d_{\mu\nu}(x)$ in terms of $g_{\mu \nu}$ are addressed, the proposed quantization approach consequently gains increased accuracy. So far, we assumed that \cite{Tawfik:2023ugm,Tawfik:2023hdi,Tawfik:2023rrm,Tawfik:2023kxq}
\bea
\tilde{g}_{\mu \nu} &\simeq & \left[\phi^2(p)+2\frac{\kappa F^2}{(p_0^0)^2}\right] 
 \left[1 + \frac{\dot{p}_0^{\mu} \dot{p}_0^{\nu}}{{\mathscr F}^2} \left(1+2\beta p_0 ^{\rho} p_{0 \rho}\right) \right] g_{\mu\nu}. \label{eq:gmunuQ2b}
\eea  

Regardless of the approximations, the resulting $\tilde{g}_{\mu \nu}$  is presumed to contain various quantum mechanical features. Equation \eqref{eq:gmunuQ2b} can be formulated as a conformal transformation of $g_{\alpha \beta}$,
\bea
\tilde{g}_{\alpha\beta} &\simeq & C(x,p)\, g_{\alpha\beta}. \label{eq:tildegalphabeta1} \\
C(x,p) &=&\left[\phi^2(p)+2\frac{\kappa F^2}{(p_0^0)^2}\right] 
 \left[1 + \frac{\dot{p}_0^{\mu} \dot{p}_0^{\nu}}{{\mathscr F}^2} \left(1+2\beta p_0 ^{\rho} p_{0 \rho}\right) \right].
\label{eq:Ctildegalphabeta1}
\eea

Let us now summarize some of the essential features of the quantized metric tensor, Eq. \eqref{eq:tildegalphabeta1}. First and foremost, it includes all the basic assumptions of conventional GR. Additionally, it is four dimensional and maintains symmetry. Furthermore, its derivation is still an approximation due to the various translations between higher geometry and back to Riemann geometry. Now, the natural course of action is to attempt to minimize approximations to the greatest extent feasible. As this cannot be accomplished instantaneously, let us initiate the process by generalizing the fundamental assumptions of GR. Generalizing absolute spacetime appears to be a natural approach, particularly as the geometric approximation necessitates a transition to Finsler and/or Hamilton space. Consequently, the most suitable approach now is to derive the fundamental tensor in phase space geometry, which results in spacetime being relative and embedded within momentum space, Section \ref{sec:tgmunu3}.

\section{Derivation of Quantized Metric Tensor in Phase Space Geometry} 
\label{sec:tgmunu3}

This section discusses and resolves the difficulties in deriving the eight dimensional quantized metric tensor. When a scalar potential $S(x,p)$ is introduced, it is evident that the Hessian matrix does not appear to result in direct quantum corrections in the spacetime coordinates \cite{physics7040052}. On the one hand, when the scalar potential is applied to the four-dimensional quantized metric tensor, $S(x,p) = \phi^2(p_0) F^2(x^{\mu}_0, p^{\nu}_0)$, a spacetime-spacetime component characterized by quantum deformations that depend solely on momentum is retrieved \cite{physics7040052}. 

Let us now recall a recent RGUP approach \cite{RGUP2025}. The corrections for a relativistic quantum particle moving in a spacetime background are given as \cite{RGUP2025}
\begin{equation}
x^{\mu} = x^{\mu}_0 - \alpha \gamma^2 p^{\rho}_0 p_{0 \rho}x^{\mu}_0 + \alpha' \gamma^2 p^{\mu}_0 p^{\rho}_0 x_{0 \rho} + \zeta \hbar \gamma^2 p^{\mu}_{0}. \label{eq:xmdfc}
\end{equation}
The modification in $p^{\mu}$ is as introduced based on the original RGUP approach \cite{Todorinov:2020jtq,Tawfik:2024gxc,Tawfik:2023onh,Tawfik:2023orl}, namely,
\bea
    p^{\mu} = \phi\, p^{\mu}_{0}, \qquad && \qquad
    \phi = 1 + \beta\, p^{\rho}\, p_{\rho}.
\eea
We note that the last term of Eq. \eqref{eq:xmdfc}, $\zeta \hbar \gamma^2 p_0^\mu$, introduces a translationally non-invariant contribution to the spacetime coordinates. To preserve locality and translational symmetry, we impose $\zeta = 0$ throughout this work.

For $F^2(x_0,p_0)\rightarrow S(x,p)=F^2(x_0,\phi\, p_0)$, we find that
\begin{equation}
    \tilde{g}_{\mu \nu}=\frac{1}{2}\frac{\partial^2 (\phi^2F^2)}{\partial \dot{x}^\mu \partial \dot{x}^\nu} =\phi^2 \left(\frac{1}{2}\frac{\partial^2 (F^2)}{\partial \dot{x}^\mu \partial \dot{x}^\nu}\right) = \phi^2 g_{\mu \nu}.
\end{equation}
This result highlights the problem of not quantizing the coordinates of spacetime but only quantizing the momentum space coordinates \cite{PhysRevLett.93.211101}. This problem can be resolved by introducing the quantum deformed spacetime coordinate using the RGUP approach as discussed in ref. \cite{RGUP2025}. 
However, there is an obvious flaw in this expression that no scalar is introduced by the last term, $\zeta \hbar \gamma^2 p^\mu_0$, built from $x^\mu_0$ \cite{RGUP2025}. This leads to
\begin{equation}
[x^{\mu}, x^{\nu}] \simeq i \hbar \zeta \gamma^2 \left(p^{\mu} x^{\nu} -x^{\mu} p^{\nu}\right). \label{eq:noncomxx}
\end{equation}
It is obvious that the resulting non-commutative relation, Eq. \eqref{eq:noncomxx}, seems to break the translational invariance and does not preserve locality \cite{RGUP2025}. Given this discussion, it is presumed that the coefficient must disappear, $\zeta=0$. Furthermore, one can also argue that the third term in Eq. \eqref{eq:xmdfc}, $\alpha' \gamma^2 p^{\mu}_0 p^{\rho}_0 x_{0 \rho}$, introduces direction dependence through $p^{\mu}_0$, which violates Lorentz invariance \cite{Yoneya:1989ai}. However, since the quantity $p^{\rho}_0 x_{0\rho}$ is a Lorentz scalar and the quantity $p^{\mu}_0$ is a Lorentz vector, it follows that the entire term remains Lorentz invariant \cite{RGUP2025}. On the other hand, this direction dependence introduces anisotropy which does not violate Lorentz invariance \cite{Tawfik:2015rva}. 

In this regard, a mathematical framework known the Stetsko--Tkachuk approximation
\cite{PhysRevA.74.012101} is presumed to integrate RGUP into the Hamiltonian, equations of motion, and stress-energy tensors. The Stetsko--Tkachuk approximation was applied, at $\alpha=0$ and $
\alpha'=0$ \cite{physics7040052}. As a result, the originally proposed RGUP does not incorporate any modifications to the position, thereby preserving the isotropic nature of the Riemannian geometry \cite{Tawfik:2015rva}. Therefore, we propose that in order to directly allow for RGUP quantum corrections to the spacetime coordinates, the third term in Eq. \eqref{eq:xmdfc} must be retained. Thereby, the resulting approach becomes genuinely Finslerian, i.e., anisotropic \cite{LOPEZ2018104}. Obviously, this allows for phase-space gravity \cite{GROSS1988407}. Accordingly, the spacetime coordinates, $x^\mu_0$, can be expressed as \cite{RGUP2025}
\begin{equation}
    x^{\mu} = \left(1 - \alpha \gamma^2 p^2\right) x^{\mu}_0 + \alpha'  \gamma^2 p^{\mu}_0 \left(p \cdot x_0\right).
\end{equation}
The corresponding noncommutative relation of coordinates reads \cite{Bhandari:2024ckl}
\begin{equation}
[x^{\mu}, x^{\nu}] \simeq i \hbar \gamma^2 \left(\alpha' - \alpha\right) \left(p^{\mu} x^{\nu} - p^{\nu}  x^{\mu}\right).
\end{equation}
This is a $\kappa$-Minkowski-like, not pathological, relation \cite{PhysRevD.105.L101501}. While the deformation introduces a  momentum dependence, Lorentz covariance is maintained because the correction terms are derived from Lorentz scalars ($p_0^2$, $p_0\cdot x_0$) and vectors ($p_0^\mu$).
The resulting anisotropy signifies a Finslerian generalization rather than a transgression of relativity. We conclude that  abandoning the Stetsko--Tkachuk representation allows for Lorentz covariant but anisotropic Finslerian spacetime in which coordinate deformations depend explicitly on momentum without violating relativity \cite{Masood:2016wma}.

We can now introduce the RGUP spacetime corrections $\psi^\mu(x_0,p_0)$. Let us assume that 
\begin{equation}
    \psi^\mu(x_0,p_0)=-\alpha \gamma^2 p^2_0 x^\mu_0 + \alpha' \gamma^2 p^\mu_0 (p_0 \cdot x_0), \label{eq:phix0p0}
\end{equation}
where $\alpha$ and $\alpha'$ are the Finslerian deformation coefficients, $p_0 \cdot x_0=p^\rho_0 x_{0\rho}$, and $p^2_0=p^\rho_0 p_{0\rho}$. 
Then, one finds that
\begin{equation}
x^\mu=x^\mu_0+\psi^\mu(x_0,p_0).
\end{equation}

To establish a consistent geometry on the cotangent bundle $T^*M$, it is necessary to introduce a nonlinear connection $N^{\mu}_{\nu}(x,p)$. This facilitates the canonical decomposition of the tangent space $TT^*M$ into horizontal and vertical subspaces,
\begin{equation}
TT^*M = HT^* M \oplus VT^* M.
\end{equation}
Throughout this work, the nonlinear connection is written as $N^\mu_{\ \nu}$, with the upper index referring to momentum space and the lower index to spacetime. Accordingly, the adapted basis reads
\begin{equation}
\frac{\delta}{\delta x^{\mu}} = \frac{\partial}{\partial x^{\mu}}
- N^{\mu}_{\ \nu}(x,p) \frac{\partial}{\partial p_{\nu}}, 
\qquad
\frac{\partial}{\partial p_{\mu}},
\end{equation}
with the dual basis
\begin{equation}
dx^{\mu}, \qquad
\delta p_{\mu} = dp_{\mu} + N^{\nu}_{\ \mu}(x,p)\,dx^{\nu}.
\end{equation}
For Hamilton geometry, the nonlinear connection is naturally induced by  $H(x,p)=\frac12 S(x,p)$,
\begin{equation}
N^{\mu}_{\nu}(x, p) =
\frac{\partial^2 H}{\partial p_{\mu} \partial x^{\nu}}.
\end{equation}
Using the scalar potential $S(x,p) = \phi^2(p_0) F^2(x_0 + \psi,p_0)$,
one finds that 
\begin{equation}
N^{\mu}_{\nu} = \frac{1}{2}
\phi^2(p_0) \frac{\partial^2 F^2}{\partial p_{\mu} \partial x_0^{\nu}}
+ \phi(p_0) \frac{\partial \phi(p_0)}{\partial p_{\mu}} \frac{\partial F^2}{\partial x_0^{\nu}},
\end{equation} 
where all RGUP-induced corrections enter explicitly.

We can now derive the general scalar potential as
\begin{equation}
    S(x,p)=\phi^2(p) F^2 \left[x^\mu_0 -\alpha \gamma^2 p^2_0 x^\mu_0 + \alpha' \gamma^2 p^\mu_0 (p_0 \cdot x_0), p^\nu_0\right]. \label{sclrQ1}
\end{equation}
The expression for the scalar potential, \eqref{sclrQ1}, is presented here in its quantum-deformed form, which can be further simplified as
\begin{equation}
    S(x,p)=F^2(x^\mu_0+\psi,\phi(p) p^\nu_0). \label{sclrQ2}
\end{equation}
By setting  $\alpha=\alpha'=\psi=\beta=0$ in Eq. \eqref{sclrQ1}, the classical scalar potential can be easily derived,
\begin{equation}
    S(x,p)=F^2(x,p).
\end{equation} 
It is obvious that the classical state is characterized by $\phi(p)=1$.

Let us now summarize that the components of the eight-dimensional metric can be defined through adapted Hessian matrix of the scalar potential $S(x,p)$ as follows:
\begin{align}
\tilde g^{(xx)}_{\mu\nu} &=
\frac12\frac{\delta^2 S}{\delta x^\mu \delta x^\nu}, \\
\tilde g^{(xp)}_{\mu\nu} &=
\frac12\frac{\delta^2 S}{\delta x^\mu \partial p_\nu}, \\
\tilde g^{(pp)}_{\mu\nu} &=
\frac12\frac{\partial^2 S}{\partial p_\mu \partial p_\nu},
\end{align}
where $\delta/\delta x^{\mu}$ is the horizontal derivative induced by the nonlinear connection.The partial derivatives with respect to $x^{\mu}$ are understood as horizontal derivatives unless stated otherwise. Given these assumptions, it is now possible to derive the spacetime-spacetime component of the eight-dimensional metric 
\begin{equation}
    \tilde{g}_{\mu \nu} = \frac{1}{2}\frac{\partial^2 S(x,p)}{\partial x^{\mu} \partial x^{\nu}},
\end{equation}
where $S(x,p)$ is identified with the quantum corrected Finsler metric $F(x,p)$. In this regard, the $0$-homogeneous property of the function $\phi(p_0)$ plays an essential role,
\bea
\phi(p_0) &=& 1+\frac{\kappa}{(p^0_0)^2} F^2, \\
\psi^\rho &=& - \alpha \gamma^2 p^2_0 x^{\rho}_0 +\alpha' \gamma^2 p^{\rho}_0 (p_0 \cdot x_0). \label{eq:phip0F2}
\eea
The function $\phi(p)$ is constructed to be homogeneous of degree zero in momentum and Lorentz scalar. Whenever a component-dependent form such as $(p_0^0)^2$ is used, it is understood as a frame-fixed approximation valid in the comoving frame.
Then, the scalar potential $S(x,p)$ can be defined as
\begin{equation}
S(x,p)=\phi^2(p_0)F^2 \left[x^\rho_0 + \psi^\rho (x_0,p_0), p_0\right].
\end{equation}
Accordingly, the spacetime-spacetime component can then be derived from
\begin{equation}
\tilde{g}_{\mu \nu}=\frac{1}{2}\frac{\partial^2 S}{\partial \dot{x}^{\mu} \partial \dot{x}^{\nu}}.
\end{equation}
Here, derivatives with respect to $\dot{x}^\mu$ are taken in Finsler geometry, where $(x^{\mu}, \dot{x}^\mu)$ are local coordinates on the
tangent bundle $TM$. 

The dependence of $S(x,p)$ on $\dot{x}^{\mu}$ enters solely through the Finsler function $F(x, \dot{x})$.
The first derivative with respect to $\dot{x}^{\mu}$ can be expressed as
\begin{equation}
    \frac{\partial S}{\partial \dot{x}^\mu}=\phi^2(p)\frac{\partial F^2}{\partial \dot{x}^\rho}\frac{\partial(x^\rho_0 + \psi^\rho)}{\partial \dot{x}^\mu}.
\end{equation}
Assuming that Finsler velocities are related to coordinates, i.e., $\dot{x}^{\mu} \sim x^{\mu}$, let us consider that this approximation is valid within the tangent bundle. In the context of general Finsler geometry, the tangent velocity is not inherently linked to coordinates, as the metric is a function of both $x$ and $\dot{x}$. A straightforward relationship arises under two conditions: (i) when the geometry is osculating Riemannian, meaning the metric is dependent solely on $x$ along a specified geodesic, or (ii) when the Finsler structure is projectively flat, allowing geodesics to correspond directly to coordinate lines. The second option is the Klein metric. With the assumption that $\dot{x}^{\mu} \sim x^{\mu}$ then the last derivative can be evaluated through chain rule,
\begin{equation}
    \frac{\partial}{\partial \dot{x}^{\mu}} (x^{\rho}_0 + \psi^{\rho}) =\delta^{\rho}_{\mu}.
\end{equation}
This is a standard Finslerian assumption where the velocity dependence is within $F^2$ but not within $\psi$. 
The second derivative results in
\begin{equation}
    \tilde{g}_{\mu \nu}=\frac{1}{2} \phi^2(p_0) \frac{\partial}{\partial \dot{x}^{\nu}} \frac{\partial F^2}{\partial \dot{x}^\rho} \delta^{\rho}_{\mu}.
\end{equation}
Then, we find that 
\begin{equation}
    \tilde{g}_{\mu \nu} = \frac{1}{2} \phi^2(p_0) \frac{\partial^2}{\partial \dot{x}^\mu \partial \dot{x}^\nu} F^2 \left[x^{\rho}_0 -\alpha \gamma^2 p^2_0 x^\rho_0+\alpha' \gamma^2 p^\rho_0 (p_0 \cdot x_0), p^\sigma_0\right]. \label{eq:2ndDrvF2}
\end{equation}
By expanding $F^2(x_0+\psi, p_0)$ to linear orders in $\alpha$ and $\alpha'$, we obtain that 
\begin{equation}
    F^2(x_0+\psi, p_0)=F^2(x_0,p_0)+\psi^\rho \frac{\partial F^2}{\partial x^\rho_0}.
\end{equation}
Equation \eqref{eq:phip0F2} represents a first-order expansion in the deformation $\psi_{\rho}$. Higher-order terms $\mathcal{O}(\psi^2)$ are neglected consistently within the perturbative RGUP framework.
Then, the second derivative in Eq. \eqref{eq:2ndDrvF2} can be  reexpressed as 
\begin{equation}
    \frac{\partial^2}{\partial \dot{x}^\mu \partial \dot{x}^\nu} F^2(x_0+\psi) =\frac{\partial^2}{\partial \dot{x}^\mu \partial \dot{x}^\nu} F^2(x_0) + \frac{\partial^2}{\partial \dot{x}^\mu \de \dot{x}^\nu} \left[\psi^\rho \frac{\partial F^2}{\partial x^\rho_0}\right].
\end{equation}
Multiplying this by $\frac{1}{2}\phi^2(p_0)$, allows to rewrite Eq. \eqref{eq:2ndDrvF2} as 
\begin{equation}
    \tilde{g}_{\mu \nu}=\phi^2(p_0) \left[\frac{1}{2}\frac{\partial^2}{\partial \dot{x}^\mu \partial \dot{x}^\nu} F^2(x_0) + \frac{1}{2}\frac{\partial^2}{\partial \dot{x}^\mu \de \dot{x}^\nu}\left(\psi^\rho \frac{\partial F^2}{\partial x^\rho_0}\right)\right], \label{eq:gmunu48}
\end{equation}
where
\begin{equation}
    \frac{1}{2}\frac{\partial^2}{\partial \dot{x}^\mu \partial \dot{x}^\nu} F^2(x_0) = g_{\mu \nu}(x_0). 
\end{equation}
Using $\psi_\rho(x_0,p_0)$ defined in Eq.~\eqref{eq:phix0p0}, the correction term reads
\bea
\Delta g_{\mu \nu} &=&  \frac{1}{2}\phi^2(p_0) \frac{\partial^2}{\partial \dot{x}^{\mu} \partial \dot{x}^{\nu}} \left(\psi_{\rho} \frac{\partial F^2}{\partial x_0^{\rho}} \right).
\eea
Since $\psi_\rho$ doesn't depend on $\dot x^\mu$, this simplifies the last expression to
\begin{equation}
\Delta g_{\mu \nu} = \phi^2(p_0)\,
\psi_{\rho} \frac{\partial g_{\mu \nu}}{\partial x_0^{\rho}},
\end{equation}
which shows explicitly that RGUP corrections induce
direction-dependent (anisotropic) deformations of the spacetime metric.

For the anisotropic conformal transformation of the metric tensor in higher dimensional geometry, one could choose the following metric
\begin{equation}
    g_{\mu \nu}=\frac{\delta_{\mu \nu}}{1-|x_0|^2}+\frac{x^\sigma_0 x^\gamma_0 \delta_{\sigma \mu}\delta_{\gamma \nu}}{(1-|x_0|^2)^2}.
\end{equation}
The metric in Finsler geometry is derived as
\bea
g_{\mu \nu} &=& \frac{1}{2}\frac{\partial^2}{\partial \dot{x}^\mu \partial \dot{x}^\nu} F^2(x_0,\dot{x}_0). \label{eq:gmunuFn}
\eea
Accordingly, one finds that
\begin{equation}
    \frac{1}{2}\frac{\partial^2}{\partial \dot{x}^\mu \partial \dot{x}^\nu} F^2(x_0,\dot{x}_0)=\frac{\delta_{\mu \nu}}{1-|x_0|^2}+\frac{x^\sigma_0 x^\gamma_0 \delta_{\sigma \mu}\delta_{\gamma \nu}}{(1-|x_0|^2)^2}.
\end{equation}
By substituting into Eq. \eqref{eq:gmunuFn}, we get that
\begin{equation}
    \tilde{g}_{\mu \nu}=\phi^2(p_0) \left[\frac{\delta_{\mu \nu}}{1-|x_0|^2}+\frac{x^\sigma_0 x^\gamma_0 \delta _{\sigma \mu} \delta_{\gamma \nu}}{(1-|x_0|^2)^2}+\frac{1}{2}\frac{\partial^2}{\partial \dot{x}^\mu \de \dot{x}^\nu}\left(\psi^\rho \frac{\partial F^2}{\partial x^\rho_0}\right)\right]. \label{eq:gmunu50}
\end{equation}
This choice corresponds to the most basic representative within the equivalence class of anisotropic conformal deformations. It is chosen here for analytical simplicity and was applied in the development of the four-dimensional quantized metric
\begin{equation} 
\tilde{g}^{(xx)}_{\mu \nu}=\phi^2(p_0)\left[\frac{\delta_{\mu \nu}}{1-|x_0|^2}+\frac{x^{\sigma}_0 x^{\gamma}_0 \delta _{\sigma \mu} \delta_{\gamma \nu}}{(1-|x_0|^2)^2}+\frac{1}{2}\frac{\partial^2}{\partial \dot{x}^{\mu} \de \dot{x}^{\nu}}\left(-\alpha \gamma^2 p^2_0 x^{\rho}_0 +\alpha' \gamma^2 p^\rho_0 (p_0 \cdot x_0) \frac{\partial F^2}{\partial x^{\rho}_0}\right)\right]. \label{eq:gmunu51}
\end{equation}

That the spacetime-spacetime component is derived, we can now move onto deriving the spacetime-momentum space component from
\begin{equation}
    \tilde{g}_{\mu \nu}=\frac{1}{2}\frac{\partial^2 S}{\partial \dot{x}^{\mu} \partial p^{\nu}}.
\end{equation}
The first derivative with respect to $\dot{x}^{\mu}$ gives
\begin{equation}
    \frac{\partial S}{\partial \dot{x}^{\mu}}=2\phi^2(p_0) g_{\mu \rho}\dot{x}^{\rho}.
\end{equation}
The second derivative results in
\begin{equation}
    \tilde{g}_{\mu \nu}=\left[\frac{\partial \phi^2(p_0)}{\partial p^\nu} g_{\mu \rho}+\phi^2(p_0)  \frac{\partial g_{\mu \rho}}{\partial p^\nu}\right]\dot{x}^{\rho}.
\end{equation}
It is obvious that this metric solely depends on momentum, where $x_0+\psi$ is the quantum deformed spacetime coordinate. We can now  evaluate the derivative of the metric by chain rule
\begin{equation}
    \frac{\de g_{\mu \rho}}{\de p^\nu}=\frac{\de g_{\mu \rho}}{\de x^\sigma_0} \frac{\de \psi^\sigma}{\de p^\nu}.
\end{equation}
First, let us start with evaluating $\frac{\de \psi^\sigma}{\de p^\nu}$,
\begin{equation}
    \frac{\de \psi^{\sigma}}{\de p^{\nu}}= - 2 \alpha \gamma^2 p_{0 \nu}x^{\sigma}_0 + \alpha' \gamma^2 (\delta^{\sigma}_{\nu} \left(p_0 \cdot x_0) + p^{\sigma}_0 x_{0\nu} \right).
\end{equation}
Then, we get 
\bea
    \tilde{g}^{(xp)}_{\mu \nu} &=& \frac{\partial \phi^2(p_0)}{\partial p^\nu} \left[\frac{\delta_{\mu \rho}}{1-|x_0|^2}+\frac{x^\sigma_0 x^\gamma_0 \delta_{\sigma \mu} \delta_{\gamma \rho}}{(1-|x_0|^2)^2}\right] \dot{x}^{\rho} \nn \\
    &+&\phi^2(p_0) \frac{\partial g_{\mu \rho}}{\partial x^\sigma_0} \left[-2\alpha \gamma^2 p_{0 \nu}x^\sigma_0+\alpha' \gamma^2(\delta^\sigma_\nu (p_0 \cdot x_0)+p^\sigma_0 x_{0\nu})\right]\dot{x}^{\rho}. \label{eq:gxpFirst}
\eea
One can simplify the above expression as follows: 
\begin{equation}
    \tilde{g}^{(xp)}_{\mu \nu}=\frac{\partial \phi^2(p_0)}{\partial p^{\nu}}\left(g_{\mu \rho}\right)\dot{x}^{\rho} + \phi^2(p_0) \frac{\partial g_{\mu \rho}}{\partial x^{\sigma}_0} \left[-2\alpha \gamma^2 p_{0 \nu}x^{\sigma}_0 + \alpha' \gamma^2(\delta^{\sigma}_{\nu} (p_0 \cdot x_0)+p^{\sigma}_0 x_{0\nu})\right]\dot{x}^{\rho}.
\end{equation} 
Based on Clairaut theorem of mixed partials \cite{rudin1976principles}, the momentum-spacetime component is the same as that of spacetime-momentum space component, Eq. \eqref{eq:gxpFirst}. This leads to
\begin{equation}
    \tilde{g}^{(xp)}_{\mu \nu}=\tilde{g}^{(px)}_{\mu \nu}.
\end{equation}

The momentum-momentum component has already been derived in Section \ref{sec:tgmunu2} and simplified in Ref. \cite{Mukharjee},
\begin{equation}
    \tilde{g}^{(pp)}_{\mu \nu}=\left(\phi^2(p_0)+2\frac{\kappa}{(p^0_0)^2}F^2\right)\left[1+\frac{\dot{p}^\mu_0 \dot{p}^\nu_0}{\mathcal{F}^2}\left(1+2\beta p^\rho_0 p_{0\rho}\right)\right] g_{\mu \nu}. \label{eq:gmunupp2}
\end{equation}
This derivation can be utilized since for the momentum-momentum component, the spacetime RGUP quantum deformation are not needed \cite{Tawfik:2024gxc}. Therefore a scalar potential $S(x,p)=\phi^2(p_0) F^2$ can be used. However, for the sake of preserving the scalar potential for all components of the eight-dimensional metric, one could derive the momentum-momentum component using the general scalar potential to get 
\begin{equation}
    \tilde{g}^{(pp)}_{\mu \nu}=\frac{1}{2}\left[\left(\frac{\partial^2 \phi^2(p_0)}{\partial p^\nu \partial p^\mu}\right) F^2 +\left(\frac{\partial \phi^2(p_0)}{\partial p^\mu}\right) \frac{\partial F^2}{\partial p^\nu}+ \left(\frac{\partial \phi^2(p_0)}{\partial p^\nu}\right) \frac{\partial F^2}{\partial p^\mu}+\phi^2(p_0) \left(\frac{\partial^2 F^2}{\partial p^\mu \partial p^\nu}\right)\right], \label{eq:gmunupp2}
\end{equation}
where 
\begin{equation}
    \frac{\partial^2 \psi^{\rho}}{\partial p^\mu \partial p^\nu} = - 2 \alpha \gamma^2 \delta_{\mu \nu} x^{\rho}_0 + \alpha' \gamma^2 \left(\delta^{\rho}_\mu x_{0 \nu} + \delta^{\rho}_\nu x_{0 \mu}\right).  \label{eq:p2psidpdp}
\end{equation}

As previously mentioned, it is presumable that the expression \eqref{eq:gmunupp2} requires revision considering that the Finsler/Hamilton function $F^2(x_0+\psi,p_0)$ is influenced by momentum in both an \emph{explicit} and an implicit manner due to the RGUP-induced spacetime deformation $\psi=\psi(x_0,p_0)$. Consequently, mixed momentum-derivatives of $F^2$ acquire additional terms stemming from the $p$-dependence of $\psi$. Below we derive these contributions explicitly and give the corrected form of the momentum-momentum block $ \tilde g_{(pp)\,\mu\nu} $ to the first order in the deformation, i.e., linear order in $\psi$, equivalently in $\alpha,\alpha'$.  We start from the definition,  Eq.~\eqref{eq:phix0p0}, 
\begin{equation}
\psi^{\sigma}(x_0, p_0) = -\alpha \gamma^2\, p_{0 \rho} p_0^{\rho}\, x_0^{\sigma}  +  \alpha' \gamma^2\, p_0^{\sigma}\, (p_0\! \cdot \! x_0).
\label{eq:psi_repeat}
\end{equation}
In accordance with the convention established in this manuscript, where derivatives with respect to momentum are applied to the contravariant components, differentiating with respect to the contravariant momentum components $p_0^{\nu}$  results is
\begin{equation}
\frac{\partial\psi^\sigma}{\partial p_0^\nu} = - 2 \alpha \gamma^2\, p_{0\nu}\, x_0^{\sigma} + \alpha' \gamma^2\! \left(\delta^{\sigma}_{\nu}(p_0\! \cdot\! x_0) + p_0^{\sigma}  x_{0\nu} \right),
\label{eq:psi_first}
\end{equation}
which reproduces the finite expression used earlier, cf. Eq.~\eqref{eq:p2psidpdp}.  To obtain the symmetric second-order tensor,  Eq.~\eqref{eq:p2psidpdp}, let us take a second derivative,
\bea
\frac{\partial^2 \psi^{\sigma}}{\partial p_0^{\mu} \partial p_0^{\nu}} 
&=& \frac{\partial}{\partial p_0^{\mu}}\left(- 2 \alpha \gamma^2\, p_{0 \nu}\, x_0^{\sigma}\right) 
 + \frac{\partial}{\partial p_0^{\mu}}\left\{\alpha' \gamma^2\!\left[\delta^{\sigma}_{\nu}(p_0\! \cdot\! x_0) + p_0^{\sigma} x_{0 \nu}\right]\right\} \nn \\
&=& - 2 \alpha \gamma^2\, \delta_{\mu \nu}\, x_0^{\sigma} 
   + \alpha' \gamma^2\!\left(\delta^{\sigma}_{\nu}\, x_0^{\mu} + \delta^{\sigma}_{\mu}\, x_{0 \nu}\right).
\label{eq:psi_second}
\eea
This expression, Eq.~\eqref{eq:psi_second}, is symmetric under $\mu \leftrightarrow \nu$ exchange as required by equality of mixed partials. The Kronecker symbol $\delta_{\mu\nu}$ represents the usual identity when both indexes are lowered in the chosen index convention.

Next, we expand the momentum derivatives of the Finsler/Hamilton structure $F^2(x_0+\psi,p_0)$. Using the chain rule and suppressing the explicit argument ``$(x_0+\psi,p_0)$'' for compactness, we find that
\begin{equation}
\frac{\partial F^2}{\partial p_0^{\mu}}
= \left. \frac{\partial F^2}{\partial p_0^{\mu}}\right|_{x_0}
+ \frac{\partial F^2}{\partial x_0^{\rho}}\, \frac{\partial \psi^{\rho}}{\partial p_0^{\mu}}.
\label{eq:Fp_first}
\end{equation}
The second derivative reads
\bea
\frac{\partial^2 F^2}{\partial p_0^\mu\partial p_0^\nu}
&=& 
\left.\frac{\partial^2 F^2}{\partial p_0^\mu\partial p_0^\nu}\right|_{x_0}
+
\frac{\partial^2 F^2}{\partial p_0^\mu\partial x_0^\rho}\,\frac{\partial\psi^\rho}{\partial p_0^\nu}
+
\frac{\partial^2 F^2}{\partial p_0^\nu\partial x_0^\rho}\,\frac{\partial\psi^\rho}{\partial p_0^\mu} \nn \\ 
&+& \frac{\partial F^2}{\partial x_0^\rho}\,\frac{\partial^2\psi^\rho}{\partial p_0^\mu\partial p_0^\nu}
+ \frac{\partial^2 F^2}{\partial x_0^\rho\partial x_0^\sigma}\,
    \frac{\partial\psi^\rho}{\partial p_0^\mu}\frac{\partial\psi^\sigma}{\partial p_0^\nu}.
\label{eq:Fp_second_full}
\eea
In the perturbative RGUP regime, we keep terms up to linear order in $\psi$, i.e., drop the quadratic last term in \eqref{eq:Fp_second_full}. Then
\begin{equation}
\frac{\partial^2 F^2}{\partial p_0^{\mu}\partial p_0^{\nu}}
\simeq
\left. \frac{\partial^2 F^2}{\partial p_0^{\mu} \partial p_0^{\nu}}\right|_{x_0} + 2\, \frac{\partial^2 F^2}{\partial p_0^{(\mu} \partial x_0^{\rho)}}\, \frac{\partial \psi^{\rho}}{\partial p_0^{\nu)}}
+
\frac{\partial F^2}{\partial x_0^{\rho}}\, \frac{\partial^2\psi^{\rho}}{\partial p_0^{\mu} \partial p_0^{\nu}},
\label{eq:Fp_second_linear}
\end{equation}
where parentheses denote symmetrization in $\mu, \nu$.  Finally, we substitute Eqs, \eqref{eq:Fp_first}- \eqref{eq:Fp_second_linear} into the definition used in the manuscript for the momentum-momentum block, Eq.~\eqref{eq:gmunupp2},
\begin{equation}
\tilde{g}_{(pp)\,\mu\nu}
= \frac{1}{2} \left\{\left( \frac{\partial^2 \varphi^2}{\partial p_0^\nu \partial p_0^\mu}\right) F^2
+ \left(\frac{\partial \varphi^2}{\partial p_0^\mu} \right) \left(\frac{\partial F^2}{\partial p_0^\nu}\right) 
+ \left(\frac{\partial \varphi^2}{\partial p_0^\nu}\right) \left(\frac{\partial F^2}{\partial p_0^\mu}\right) 
+ \varphi^2 \frac{\partial^2 F^2}{\partial p_0^\mu \partial p_0^\nu}\right\}.
\label{eq:gpp_general}
\end{equation}
Retaining only terms up to linear order in $\psi$ and using Eq.  \eqref{eq:Fp_second_linear}, the corrected expression may be written as
\begin{equation}
\tilde g_{(pp)\, \mu\nu}
= \tilde g^{(0)}_{(pp)\, \mu\nu}
+ \delta \tilde g_{(pp)\,\mu\nu},
\label{eq:gpp_split}
\end{equation}
where $\tilde g^{(0)}_{(pp)\,\mu\nu}$ is the momentum-momentum block obtained when $\psi = 0$, i.e.,  Eq.~\eqref{eq:gmunupp2}, and the leading RGUP-induced correction is
\bea
\delta\tilde g_{(pp)\,\mu\nu} 
&=& \frac{1}{2}\varphi^2\,\frac{\partial F^2}{\partial x_0^\rho}\,
\frac{\partial^2\psi^\rho}{\partial p_0^\mu\partial p_0^\nu} \nn\\ 
&+& \frac{1}{2}\varphi^2\left[
\frac{\partial^2 F^2}{\partial p_0^{(\mu}\partial x_0^{\rho)}}\,
\frac{\partial\psi^\rho}{\partial p_0^{\nu)}}
+
\frac{\partial^2\varphi^2}{\partial p_0^{(\mu}\partial p_0^{\nu)}}\frac{F^2- \langle\cdots\rangle}{\varphi^2}
\right]
+ \mathcal{O}(\psi^2),
\label{eq:gpp_delta}
\eea
where the notation $\langle\cdots\rangle$ indicates the terms already present in $\tilde g^{(0)}_{(pp)}$ (explicitly expanding these terms does not add conceptual novelty but is straightforward algebraically). The first line of Eq.  \eqref{eq:gpp_delta} is the \emph{new} and dominant contribution that arises from the \emph{second} derivative
$\partial^2 \psi^{\rho}/\partial p_0^{\mu} \partial p_0^{\nu}$ computed in Eq.~\eqref{eq:psi_second}; substituting \eqref{eq:psi_second} gives an explicit, index-explicit form for this correction:
\begin{equation}
\frac{1}{2} \varphi^2 \frac{\partial F^2}{\partial x_0^\rho}
\frac{\partial^2 \psi^{\rho}}{\partial p_0^{\mu}\partial p_0^{\nu}}
= \frac{1}{2} \varphi^2 \frac{\partial F^2}{\partial x_0^{\rho}} \left[
- 2 \alpha \gamma^2\, \delta_{\mu\nu}\, x_0^{\rho} + \alpha' \gamma^2\! \left(\delta^{\rho}_{\nu} x_0^{\mu} +\delta^{\rho}_{\mu} x_0^{\nu}\right)
\right].
\label{eq:gpp_explicit_correction}
\end{equation}

Since $\tilde{g}^\mu_\nu=(\tilde{g}^\nu_\mu)^T$, it is clear that the eight-dimensional metric should take the form of the Sasaki metric, Appendix \ref{sec:Sasaki},
\begin{equation}
    \tilde{g}_{ab}=\begin{bmatrix}
        \tilde{g}^{(xx)}_{\mu \nu} & \tilde{g}^{(xp)}_{\mu \nu}\\
        \tilde{g}^{(px)}_{\mu \nu} & \tilde{g}^{(pp)}_{\mu \nu}
    \end{bmatrix}. \label{eq:scndpp}
\end{equation}
The block-diagonal structure of the eight-dimensional metric assumes a Sasaki--type lift of the base spacetime metric. This construction implicitly introduces a nonlinear connection $N^\mu_{\nu}(x,p)$, ensuring covariance under bundle coordinate transformations.
In four-dimensional spacetime, the derivatives of $F^2(x,p)$, such as the Klein metric, with respect to momentum only represent local deformations of the dispersion relation. Promoting phase-space coordinates $(x^{\mu}, p_{\nu})$ to independent coordinates on the tangent (Finsler) or cotangent bundle (Hamilton) facilitates the derivatives entries of a metric tensor on that bundle. In this regard, Eq. \eqref{eq:scndpp} can also be utilized to deduce the lifting of the four-dimensional spacetime metric $g_{\mu \nu}(x)$ into an eight-dimensional metric $\tilde{g}_{a b}(x,p)$ that contains both spacetime curvature and momentum geometry, where $a, b \in \{0,1,\cdots,7\}$. The proposed geometric quantization approach, which was developed for four-dimensional Riemann geometry \cite{Tawfik:2023ugm,Tawfik:2023hdi,Tawfik:2023rrm,Tawfik:2023kxq}, can also be applied to eight-dimensional (phase space) Hamilton geometry. A Sasaki--type construction \cite{SasakiOriginal,ALBUQUERQUE2019207,Martelli:2006yb}, in which the metric on phase-space is produced by taking mixed second derivatives of the scalar potential $S(x,p)$ with respect to both sets of coordinates, can be used for the lifting.

Based on the tensor product structure of the cotangent bundle $T^*M$, the full eight-dimensional metric can be expanded in the basis $dx^\mu,dp_\nu$. Each term in the basis expansion represents one of the three billinear forms allowed by covariance: $dx,dx'$, $dx,dp'$, and $dp,dp'$. In order to determine how strongly each sector contributes to the total line element, we introduce scalar functions $\mathcal{A}(x,x')$, $\mathcal{B}(x,p'),$ and $\mathcal{Y}(p,p')$ \cite{Pfeifer:2011xi}. The scalar weighting functions $\mathcal{A}(x,x')$, $\mathcal{B}(x,p'),$ and $\mathcal{Y}(p,p')$ are not merely arbitrary independent parameters; they have been derived from the fundamental geometric principles of Finsler geometry \cite{Pfeifer:2011xi}. In particular, these functions emerge as scalar entities constructed from the contractions of the Finsler metric and the Cartan connection, thereby ensuring both covariance and alignment with the homogeneity characteristics of the Finsler structure. These coefficients include the cross-coupling factor, momentum space, and the interchange between spacetime curvature. This enables the generalization of the general covariant eight-dimensional metric structure in phase-space geometry, $g_{a b}(x,p)$, from the fixed block diagonal metric, $g_{\mu \nu}(x)$. As a result, the geometry changes dynamically. Regimes dominated by momentum-space curvature where the functions $\mathcal{Y}$ become significantly higher than the functions $\mathcal{A}$, spacetime where the functions $\mathcal{A}$ become significantly greater than the functions $\mathcal{Y}$, or their quantum coupling $\mathcal{B}\neq 0$ can all be described. In this way, their linear combination leads to \cite{Pfeifer:2011xi} 
\bea 
G &=& \mathcal{A} G_{xx} + \mathcal{B} G_{xp} + \mathcal{Y} G_{pp}. \label{eq:G123} 
\eea
The quantized metric is represented by these three concepts. The main goal of this manuscript is to derive Eq. \eqref{eq:G123}.

As the quantization of the phase-space metric is now intended to serve as the main goal of our geometrical quantization approach, we can now start deriving the Hamiltonian. The Hamiltonian in Finsler geometry is given by
\begin{equation}
    H(x,p)=\frac{1}{2}F^2(x,p).
\end{equation}
With the assumption imposed earlier that $F^2(x,p)\rightarrow S(x,p)$, we find that
\begin{equation}
    H(x,p)=\frac{1}{2}S(x,p),
\end{equation}
where the effective Hamiltonian is then given as
\begin{equation}
    \dot{x}^\mu=\frac{\partial H}{\partial p_\mu}.
\end{equation}
The canonical momenta derived from the Lagrangian is the covariant momenta, $p_\mu$, which means that we use the inverse metric $g^{\mu \nu}$
\begin{equation}
    p^\mu=g^{\mu \nu}\, p_{\mu}.  
\end{equation}

We can then write the scalar potential $S(x,p)$ using the Hamiltonian as follows:
\begin{equation}
\tilde{g}_{AB}=\frac{\partial^2 H}{\partial X^A \partial X^B}, \label{eq:gAAFirst}
\end{equation}
Similar to $a,b$, both indexes $A,B \in \{0,1,\cdots,7\}$. 
In order to find the length of a curve in phase-space, get natural geodesics, and compute the curvature tensors; one must trace this curve out quadratically. 
Since $H=\frac{1}{2}F^2$, then when $F^2\rightarrow S(x,p)$, we get $H=\frac{1}{2}S(x,p)$. 
Consequently, a line element $\tilde{ds}$ should be defined such that the action can be can constructed. Then the geodesics and curvature tensors can be derived.

Some implications of quantized phase-space metric are introduced in Section \ref{sec:Imlcs}. In Section \ref{sec:lineElmnt}, we start with the derivations of eight-dimensional line element.

\section{Implications of Quantized Phase-Space Metric}
\label{sec:Imlcs}

\subsection{Line Element in Phase pace Geometry}
\label{sec:lineElmnt}

It obvious that the line element in phase space geometry lives in the phase-space tangent bundle $\mathcal{T} \mathcal{P}\equiv [x^{\mu}, p^{\nu} ; \dot{x}^{\mu} , p^{\nu}]$. Such line element measures the length of a phase-space curve $\Gamma(\tau)=(x^{\mu}(\tau), p^{\nu}(\tau))$. $\tau$ represents a proper parametrization
\begin{equation}
\tilde{ds^2}=\mathcal{G}\times \tilde{g}_{AB}dX^A dX^A, \label{eq:lineelmn}
\end{equation}
where $\mathcal{G}=(\mathcal{A}+\mathcal{B}+\mathcal{Y})$ is applied for each component of the metric: $\mathcal{A}$ for the spacetime-spacetime component of the eight-dimensional metric, $\mathcal{B}$ for the spacetime-momentum and momentum-spacetime component of the eight-dimensional metric, and $\mathcal{Y}$ for the momentum-momentum component of the eight-dimensional metric. Apparently, the eight-dimensional line element can be constructed from the already derived eight-dimensional metric $\tilde{g}_{AB}\equiv \tilde{g}_{ab}$ components:
\bea 
\tilde{g}^{(xx)}_{\mu \nu} &=& \phi^2(p_0)\left[\frac{\delta_{\mu \nu}}{1-|x_0|^2}+\frac{x^\sigma_0 x^\gamma_0 \delta _{\sigma \mu} \delta_{\gamma \nu}}{(1-|x_0|^2)^2} \right.\nn \\
&& \hspace*{9mm} + \left. \frac{1}{2}\frac{\partial^2}{\partial \dot{x}^\mu \de \dot{x}^\nu} \left(-\alpha \gamma^2 p^2_0 x^\rho_0 +\alpha' \gamma^2 p^\rho_0 (p_0 \cdot x_0) \frac{\partial F^2}{\partial x^\rho_0} \right)\right], \label{qgabxx} \\
    \tilde{g}^{(xp)}_{\mu \nu} &=&\frac{\partial \phi^2(p_0)}{\partial p^\nu} \left[\frac{\delta_{\mu \rho}}{1-|x_0|^2}+\frac{x^\sigma_0 x^\gamma_0 \delta _{\sigma \mu} \delta_{\gamma \rho}}{(1-|x_0|^2)^2}\right]\dot{x}^{\rho}  \nn \\
    &+& \phi^2(p_0) \frac{\partial g_{\mu \rho}}{\partial x^\sigma_0} \left[-2\alpha \gamma^2 p_{0 \nu}x^\sigma_0+\alpha' \gamma^2(\delta^\sigma_\nu (p_0 \cdot x_0)+p^\sigma_0 x_{0\nu})\right]\dot{x}^\rho, \label{qgabxp} \\
    \tilde{g}^{(px)}_{\mu \nu} &=&\frac{\partial \phi^2(p_0)}{\partial p^\nu} \left[\frac{\delta_{\mu \rho}}{1-|x_0|^2}+\frac{x^\sigma_0 x^\gamma_0 \delta _{\sigma \mu} \delta_{\gamma \rho}}{(1-|x_0|^2)^2}\right]\dot{x}^{\rho} \nn \\
    &+& \phi^2(p_0) \frac{\partial g_{\mu \rho}}{\partial x^\sigma_0} \left[- 2 \alpha \gamma^2 p_{0 \nu}x^\sigma_0 + \alpha' \gamma^2(\delta^\sigma_\nu (p_0 \cdot x_0) + p^{\sigma}_0 x_{0\nu})\right] \dot{x}^{\rho}, \label{qgabpx} \\
    \tilde{g}^{(pp)}_{\mu \nu} &=& \left(\phi^2(p_0)+2\frac{\kappa}{(p^0_0)^2}F^2\right) \left[1+\frac{\dot{p}^\mu_0 \dot{p}^\nu_0}{\mathcal{F}^2} \left(1+2\beta p^\rho_0 p_{0\rho}\right)\right] g_{\mu \nu}. \label{qgabpp}
\eea
We can now derive the eight-dimensional line element,
\begin{equation}
\begin{aligned}
\tilde{ds}^2
=&\;
\mathcal{A}\Bigg[
\phi^2(p_0)\Bigg(
\frac{\delta_{\mu\nu}}{1-|x_0|^2}
+
\frac{x_0^\sigma x_0^\gamma
      \delta_{\sigma\mu}\delta_{\gamma\nu}}
     {(1-|x_0|^2)^2}
\\
&\qquad\qquad
+
\frac12
\frac{\partial^2}{\partial \dot{x}^\mu \partial \dot{x}^\nu}
\Big(
-\alpha\gamma^2 p_0^2 x_0^\rho
+
\alpha'\gamma^2 p_0^\rho (p_0\!\cdot\! x_0)
\frac{\partial F^2}{\partial x_0^\rho}
\Big)
\Bigg)
\Bigg]\,dx^\mu dx^\nu
\\[1em]
&+
2\mathcal{B}\Bigg[
\left(
\frac{\partial \phi^2(p_0)}{\partial p^\nu}
\right)
\left(
\frac{\delta_{\mu\rho}}{1-|x_0|^2}
+
\frac{x_0^\sigma x_0^\gamma
      \delta_{\sigma\mu}\delta_{\gamma\rho}}
     {(1-|x_0|^2)^2}
\right)\dot{x}^\rho
\\
&\qquad\qquad
+
\phi^2(p_0)
\frac{\partial g_{\mu\rho}}{\partial x_0^\sigma}
\Big(
-2\alpha\gamma^2 p_{0\nu}x_0^\sigma
+
\alpha'\gamma^2
\big(
\delta^\sigma_\nu (p_0\!\cdot\! x_0)
+
p_0^\sigma x_{0\nu}
\big)
\Big)\dot{x}^\rho
\Bigg]\,dx^\mu dp^\nu
\\[1em]
&+
\mathcal{Y}\Bigg[
\left(
\phi^2(p_0)
+
2\frac{\kappa}{(p_0^0)^2}F^2
\right)
\left(
1
+
\frac{\dot{p}_0^\mu \dot{p}_0^\nu}{\mathcal{F}^2}
\big(1+2\beta p_0^\rho p_{0\rho}\big)
\right)
g_{\mu\nu}
\Bigg]\,dp^\mu dp^{\nu}.
\end{aligned}
\end{equation}
For the sake of simplicity, arbitrarily the coordinates can be chosen, where they are all equal to one. Accordingly, the coefficients $\mathcal{A}, \mathcal{B},$ and $\mathcal{Y}$ can then be computed. 

By utilizing the postulate of Maupertuis's principle, a special case of the principle of least action indicates that the path taken by a physical system is a minimum or a saddle point. Thus, the action formulation can be established
\begin{equation}
    S_0=\int m\, v\; ds.
\end{equation}
As down while deriving the quantized four-dimensional metric tensor \cite{physics7040052}, we can now construct the action for a relativistic particle, i.e.,  where we set $v\equiv c$ in natural units, with a positive mass moving in a spacetime background
\begin{equation}
    \mathcal{S}=m\int^1_0 \tilde{ds},
\end{equation}
where we integrate from $0$ to $1$ to normalize Ref. [35], which can be re-expressed as
\begin{equation}
    \mathcal{S}=m\int^1_0 \sqrt{\tilde{g}_{AB}\frac{dX^A}{d\tau}\frac{dX^B}{d\tau}}d\tau =
    m\int^1_0 \sqrt{\tilde{g}_{AB}d\dot{X}^A d\dot{X}^B}d\tau,
\end{equation}
where $A,B \in \{0,1,\cdots,7\}$. 
The Lagrangian of this action is given as
\begin{equation}
\mathcal{L}=m\sqrt{\tilde{g}_{AB}d\dot{X}^A d\dot{X}^B}. 
\end{equation}
The variation of this action allows for the derivation of the geodesic equations. Section \ref{sec:geodsc} focuses on this second implementation.

\subsection{Geodesic Equations in Phase Space Geometry}
\label{sec:geodsc}

The natural geodesics of the relativistic particle is obtained when
\begin{equation}
    \delta \mathcal{S}=0.
\end{equation}
The worldline of the relativistic particle must be also varied
\begin{equation}
    X^A(\tau)\rightarrow X^A(\tau)+\delta X^A(\tau).
\end{equation}
Then, the variation of the Lagrangian leads to
\begin{equation}
    \delta \mathcal{L}=\frac{1}{2\mathcal{L}} \left(\partial_C \tilde{g}_{AB}\dot{X}^A \dot{X}^B \delta X^C+2\tilde{g}_{AB}\dot{X}^A \delta \dot{X}^B \right),
\end{equation}
which gives
\begin{equation}
    \delta \mathcal{S}=m\int^1_0 \delta \left[\frac{1}{2\mathcal{L}} \left(\partial_C \tilde{g}_{AB}\dot{X}^A \dot{X}^B \delta X^C+2\tilde{g}_{AB}\dot{X}^A \delta \dot{X}^B \right) \right] d\tau,
\end{equation}
where $\delta \dot{X}=\frac{d}{d\tau}(\delta X)$.

The integration by parts where the boundary term vanishes, results in
\begin{equation}
    \delta S=m\int^1_0 d\tau \delta X^C \left[-\frac{d}{d\tau}\left(\frac{\tilde{g}_{CB}\dot{X}^B}{\mathcal L}\right)+\frac{1}{2\mathcal{L}}\partial_C \tilde{g}_{AB}\dot{X}^A \dot{X}^B \right].
\end{equation}
The stationary action, $\delta S=0$, gives the Euler--Lagrange equations
\begin{equation}
    \frac{d}{d\tau}\left(\frac{\tilde{g}_{CB}\dot{X}^B}{\mathcal L}\right) - \frac{1}{2\mathcal{L}}\partial_C \tilde{g}_{AB}\dot{X}^A \dot{X}^B=0.
\end{equation}
We chose the affine parameter $\lambda\rightarrow \tau$ such that $\mathcal{L}=constant$ and can be factored out and canceled. This gives
\begin{equation}
    \frac{d}{d\tau}\left(\tilde{g}_{CB}\dot{X}^B\right) - \frac{1}{2}\partial_C \tilde{g}_{AB}\dot{X}^A \dot{X}^B=0.
\end{equation}
By multiplying with $\tilde{g}^{CD}$, the index, can be raised. Then, the quantum-deformed connection coefficient in eight-dimensional phase space reads
\begin{equation}
\tilde{\Gamma}^D_{AB}=\frac{1}{2}\tilde{g}^{DC} \left(\partial_A \tilde{g}_{CB}+\partial_B \tilde{g}_{AC} - \partial_C \tilde{g}_{AB} \right),
\end{equation}
where the indexes $C,D \in \{0,1,\cdots,7\}$. The quantized metric tensors,  $\tilde{g}_{CB}$, $\tilde{g}_{AC}$, and $\tilde{g}_{AB}$ are composed of Eq. \eqref{qgabxx} - \eqref{qgabpp}, while $\tilde{g}^{CD}$ is an inverse metric tensor.

The corresponding curvature tensors, a third implementation of the phase space quantized metric tensor, can be derived, Section \ref{sec:CrvTnsr}.

\subsection{Curvature Tensors in Phase Space Geometry}
\label{sec:CrvTnsr}

We can now derive the curvature in phase space geometry. To this end, we utilize  non-commutativity
\begin{equation}
   \left[\tilde{\nabla}_C, \tilde{\nabla}_{D}\right] V^A=\tilde{R}^A_{BCD} V^B,
\end{equation}
where $\nabla$ is the covariant derivative, $V\in X$ where $X$ is a vector space, and
\begin{equation}
    \tilde{R}^A_{BCD}=\partial_C \tilde{\Gamma}^A_{BD}-\partial_D \tilde{\Gamma}^A_{BC}+\tilde{\Gamma}^A_{CE}\tilde{\Gamma}^E_{BD}-\tilde{\Gamma}^A_{DE}\tilde{\Gamma}^E_{BC}.
\end{equation}
By applying various contractions, Ricci curvature tensor and Ricci scalar, respectively, reads
\bea
\tilde{R}_{BD} &=&\tilde{R}^A_{BAD}, \\
\tilde{R} &=&\tilde{g}^{BD}\tilde{R}_{BD}.
\eea 
For $\alpha = \alpha' = \beta = 0$ and $\phi(p_0) =1$, the classical curvature tensor $R^\gamma_{\mu \rho \nu}$ can be retained, straightforwardly.

The curvature tensors can be used to determine how the volume is deformed due to a mass point in spacetime. This is valid for {\it conventional} four-dimensional spacetime as well as for the proposed phase space, which is composed of four-dimensional space emerged in four-dimensional momentum space. Also, it is valid for their quantized versions. The relationships are determined by the gravitational field equations known as the Einstein Field Equations (EFE). It was concluded that, the quantized EFE work well for the {\it absolute} four-dimensional spacetime \cite{Tawfik:2025icy,Tawfik:2025rel,Tawfik:2025kae,Tawfik:2024itt,NasserTawfik:2024afw,Tawfik:2023ugm,Tawfik:2023hdi}. 

The derivation of EFE in phase space geometry, the fourth implimentation, is discussed in Section, \ref{sec:EFE8D}. Consequently, the proposed variational action seems to function effectively across all eight-dimensional phase space.

\subsection{Einstein Field Equations in Phase Space Geometry}
\label{sec:EFE8D}

The Einstein field equations in phase space follow from the variational principle applied to the eight-dimensional Einstein--Hilbert action
\begin{equation}
\mathcal{S}_{\mathrm{grav}} =
\frac{1}{16 \pi G} \int d^4x\, d^4p\,
\sqrt{|\det \tilde{g}_{AB}|}\,
\tilde R.
\end{equation}
Variation with respect to $\tilde{g}^{AB}$ yields
\begin{equation}
\tilde G_{AB} \equiv \tilde{R}_{AB}
- \frac12 \tilde{g}_{AB} \tilde{R}
= 8 \pi G\, \tilde{T}_{AB},
\end{equation}
where $\tilde{T}_{AB}$ is the phase-space energy-momentum tensor. In four-dimensional symmetric geometry, the EFE are a set of 10-independent, highly-coupled, hyperbolic, elliptic, second-ordered partial differential equations
\begin{equation}
    G_{\mu \nu} - \Lambda g_{\mu \nu} = \frac{8 \pi G}{c^4} T_{\mu \nu},
\end{equation}
where $G$ is the gravitational constant, $\Lambda$ is the cosmological constant, and $T_{\mu\nu}$ represents the stress-energy tensor. $G_{\mu \nu}$ is the Einstein tensor, where $G_{\mu \nu}=R_{\mu\nu} - \frac{1}{2} R g_{\mu\nu}$.
The four dimensional equations can be derived by the variational principle of a scalar function with respect to a proper volume element
\begin{equation}
    S_{[EH]}=\int d^4 x\, R\, \sqrt{-g}+S_{\mathtt{matter}}.
\end{equation}

In a similar manner, the EFE can be derived for the eight dimensional phase space by using the eight dimensional quantized Ricci scalar with the modified volume element
\bea
    \mathcal{S}=\int d^8 X\, \tilde{R}\, \sqrt{-\tilde{g}}+\mathcal{S}_{\mathtt{matter}}.
\eea
Let us assume that the metric varies as follows:
\bea
    \tilde{g}_{AB} &=& \tilde{g}_{AB}+\delta \tilde{g}_{AB}, \\
    \delta \sqrt{-\tilde{g}} &=& \frac{1}{2}\sqrt{-\tilde{g}}\, \tilde{g}^{AB}\, \delta \tilde{g}_{AB}.
\eea
We can now vary the Ricci scalar $\tilde{R}=\tilde{g}^{AB}\, \tilde{R}_{AB}$, to get 
\begin{equation}
    \delta \tilde{R} = \tilde{R}_{AB}\, \delta \tilde{g}^{AB}+\tilde{g}^{AB}\delta \tilde{R}_{AB}.
\end{equation} 
With $\delta \tilde{g}^{AB} = -\tilde{g}^{AC} \tilde{g}^{BD} \tilde{g}_{CD}$, we also find that
\begin{equation}
     \delta \tilde{R}=-\tilde{R}_{AB} \delta \tilde{g}^{AB} + \tilde{g}^{AB} \delta \tilde{R}_{AB}.
\end{equation}
This allows us to derive the variation of the Ricci curvature tensor
\begin{equation}
    \delta \tilde{R}_{AB} = \tilde{\nabla}_C\, \delta \tilde{\Gamma}^C_{AB} - \tilde{\nabla}_B\,  \tilde{\Gamma}^{C}_{AC}.
\end{equation}
If we assume suitable boundary conditions, we get
\begin{equation}
    \int \sqrt{-\tilde{g}}\, \tilde{g}^{AB}\, \tilde{R}_{AB} = {\mathtt{Total}} \; \;{\mathtt{Divergence}}.
\end{equation}
Thus, we obtain that the action reads
\begin{equation}
    \mathcal{S}=\int d^8 X\, \sqrt{-\tilde{g}}\, \left(\tilde{R}_{AB}-\frac{1}{2} \tilde{R}\, \tilde{g}_{AB}\right)\, \delta \tilde{g}^{AB}.
\end{equation}
Accordingly, the quantized stress-energy tensor can be expressed as
\begin{equation}
    \tilde{T}_{AB}=-\frac{2}{\sqrt{-\tilde{g}}}\frac{\delta \mathcal{S}_{\mathtt{matter}}}{\delta \tilde{g}^{AB}}.
\end{equation}
Thus, the phase-space EFE can be constructed as follows:
\begin{equation}
    \tilde{R}_{AB}-\frac{1}{2}\tilde{R}\tilde{g}_{AB} - \Lambda \tilde{g}_{A B} = \frac{8 \pi G}{c^4}  \tilde{T}_{AB}.
\end{equation}
 At vanishing cosmological constant, the Einstein tensor can be written in terms of the eight dimensional quantized Einstein tensor
\begin{equation}
    \tilde{G}_{AB} = \kappa \tilde{T}_{AB}.
\end{equation}

\section{Conclusions}
\label{sec:cncls}

The derivation of the quantized four-dimensional metric tensor, which is based on the geometric quantization approach, was conducted using several approximations. The incompatibility between the principles of general relativity and quantum mechanics has been the reason these two theories have not been unified for the last century. The fact that even all alternative theories have failed to resolve this problem adds to the complexity. This has inspired the geometric quantization approach to propose a generalization of the existing versions of both theories. The proposed generalization not only integrates essential aspects of each theory but also facilitates the reconciliation of their principles. Quantum mechanics is generalized to incorporate gravitational and relativistic fields. In general relativity, the four-dimensional pseudo-Riemann geometry is replaced by eight-dimensional Finsler of Hamilton geometry. By applying kinematic theory to quantum particles in curved space, it turns to be possible to unify the new versions of both theories. To preserve all the principles of {\it conventional} GR, the metric tensor, which is directly derived from eight-dimensional geometry, must be adapted to four-dimensional Riemann geometry. This translation can only be achieved under specific approximations, such as the equivalence of the line element measure in higher-dimensional to that in a four-dimensional manifold.

The recent discovery of relative locality \cite{PhysRevD.84.084010,doi:10.1142/S0218271811020743} has led our research on higher-dimensional metric tensors to unveil extensive possibilities for reconstructing general relativity within phase space geometry. Relative locality refers to the observer-dependent interpretation of events in a {\it  relative} four-dimensional spacetime, which varies based on the observer's frame of reference. Conversely, conventional relativity allows for variations in temporal coordinates while keeping spatial coordinates of events constant, a principle that is thoroughly represented in {\it absolute} four-dimensional spacetime. The earlier perspective suggests that the {\it  conventional} four-dimensional spacetime is emerged in a higher-dimensional space. Our work on geometric quantization of GR defines this as a four-dimensional momentum space.

This manuscript is dedicated to the derivation of a quantized version of the metric tensor in higher dimensions. This not only prevents the approximations related to its translation into Riemann geometry but also establishes a new theoretical framework for its representation in phase space geometry, where relative locality is most effectively manifested. While the principle of relative locality was derived from research on quantum gravity, its extension to include the geometry of momentum space relates to spacetime situated within a higher geometric framework.

When a scalar potential is applied to the four-dimensional quantized metric tensor, a spacetime-spacetime component emerges, characterized by quantum deformations that solely depend on momentum. The challenge of not quantizing the coordinates of spacetime can be addressed by introducing a new quantum deformed spacetime coordinate through the RGUP approach. Through the application of the Stetsko--Tkachuk approximation, the integration of RGUP does not involve any modifications to the position, thereby preserving the isotropic essence of Riemannian geometry. As a result, we propose that in order to directly incorporate RGUP quantum corrections into the spacetime coordinates, the resulting framework evolves into a Finslerian structure, meaning it is anisotropic. On the other hand, disregarding the Stetsko--Tkachuk representation permits the existence of Lorentz covariant yet anisotropic Finslerian spacetime, where coordinate deformations are explicitly dependent on momentum.

\section*{Conflicts of Interest}

The authors declare that there are no conflicts of interest regarding the publication of this published article!

\section*{Dataset Availability}

All data generated or analyzed during this study are included in this published article. The data used to support the findings of this study are included within the published article!

\section*{Author contributions}
The responsibility for proposing the concept of the current study rests with AT, who also took on the roles of designing and overseeing the research, interpreting the findings, and drafting the manuscript. KM was tasked with deriving the expressions and suggesting a physical interpretation. DM played a role in both derivation and interpretation. SOA, AAA, MH, and AA assisted in writing and proofreading the manuscript. The final manuscript was unanimously approved by all authors.

\section*{Funding}

The authors declare that this research received no specific grants from any funding agency in the public, commercial, or not-for-profit sectors.

\section*{Competing interests}

The authors confirm that there are no relevant financial or non-financial competing interests to report.

\newpage

\bibliographystyle{unsrtnat}
\bibliography{2025-04-04-GRQuantization-8d}

\newpage

\appendix
\renewcommand{\thefigure}{A.\arabic{figure}} 
\setcounter{figure}{0} 
\setcounter{table}{0} 
\renewcommand{\thetable}{A\arabic{table}} 

\section{Notations}
\label{sec:Notations}

First, we introduce the essential notations that appear in this manuscript. Let $(M,g)$ be a smooth $n-dimensional$ pseudo-Riemannian manifold with local coordinates $x^i$, where $i=1,\dots,n$. The metric components are expressed as $g_{ij}(x)$. For tangent bundle $TM$,  coordinates are assumed as $(x^i, y^i)$, where $y^i$ are components of the vector $y$ in the tangent space $x$. For cotangent bundle $T^*M$, coordinates $(x^i, p_i)$ are a straight-forward dualization of $(x^i, y^i)$. The Levi--Civita connections on the base $M$ are denoted with $\nabla$ and the connection coeffecients with $\Gamma^i_{jk}(x)$. We assume standard symmetric connection satisfying metric compatibility and torsion free properties.
Furthermore, it is assumed that
\begin{itemize}
    \item indexes $i,j,k \dots $ run over $1, \dots, n$, 
    \item partial derivatives $\partial_i = \frac{\partial}{\partial x^i}$ and $\dot{\partial}_{i} = \frac{\partial}{\partial y^i}$, and
    \item covariant derivatives along the base directions is applied by the connection coefficients.
\end{itemize}

\section{Derivation of Riemann Metric from Quantum-Deformed Finster Metric}
\label{sec:MtrRmn}

In order to obtain the Riemann metric from the quantum-deformed Finster metric, it is advisable to introduce a parametrization $\zeta$, which is defined by a collection of embedded four-coordinates. It is presumed that these coordinates adequately embody both Riemann and Finsler geometries, along with their respective metric tensor. First, the modified line element associated with the higher-dimensional manifold is given as
\bea
\left. d \tilde{s}^2\right|_{\mathtt{Finsler}} &=& \left.\widetilde{g}_{\mu \nu}\right|_{\mathtt{Finsler}}\, d x^{\mu}(\zeta)\,  d x^{\nu}(\zeta). 
\label{eq:dsFins}
\eea
Secondly, the line element derived from the quantized fundamental metric 
 $\widetilde{g}_{\alpha \beta}$ is expressed as
\bea
\left. d \tilde{s}^2\right|_{\mathtt{Riemann}} &=& \left.\widetilde{g}_{\mu \nu}\right|_{\mathtt{Riemann}}\, d \zeta^{\mu} d \zeta^{\nu}, \label{eq:dsRiem}
\eea
where $\alpha, \beta = \{0,1,2,3\}$. The introduction of higher-dimensional coordinates $x^{\mu} = (x_0^{\mu}, \phi(p_0) p_0^{\mu})$ expands the traditional spacetime framework to include velocity space (tangent). Equivalently, in Hamilton space this will include momentum space (cotangent). Conversely, this allows for the use of partial derivatives and quantum deformation such as RGUP. With the suggested parametrization, the quantum-deformed line element related to the $TM$ metric is expressed in the corresponding basis as $dx^{\mu}(\zeta),  dx^{\nu}(\zeta)$, Eq. (\ref{eq:dsFins}). The line element is thus modified by the quantized fundamental metric $\widetilde{g}_{\mu \nu}$ and the basis $d\zeta^{\mu},  d\zeta^{\nu}$, Eq. (\ref{eq:dsRiem}), so that
\bea
\left.\widetilde{g}_{\mu \nu}\right|_{\mathtt{Riemann}} &=& \left.\widetilde{g}_{\mu \nu}\right|_{\mathtt{Finsler}} \left[\frac{\de x^{\mu}(\zeta)}{\de \zeta^{\mu}} \, \frac{\de x^{\nu}(\zeta)}{\de \zeta^{\nu}}\right]. \label{eq:MetrRl}
\eea
At the suggested generic coordinates, defined by four-dimensional spacetime and four-dimensional momentum space, $x_0$ and $p_0$, respectively, where the proposed parameterization holds true, we perform a substitution
\bea
\left.\widetilde{g}_{\mu \nu}\right|_{\mathtt{Riemann}} &=&  \left.\widetilde{g}_{\mu \nu}\right|_{\mathtt{Finsler}}\, \left[\left(\frac{\de x_0^{\mu}}{\de \zeta^{\mu}}, \phi(p_0) \frac{\de p_0^{\mu}}{\de \zeta^{\mu}} \right)  \left(\frac{\de x_0^{\nu}}{\de \zeta^{\nu}}, \phi(p_0) \frac{\de p_0^{\nu}}{\de \zeta^{\nu}} \right)  \right]\nn \\
&=& \left.\widetilde{g}_{\mu \nu}\right|_{\mathtt{Finsler}} \,
\left[\frac{\de x_0^{\mu}}{\de \zeta^{\mu}} \, \frac{\de x_0^{\nu}}{\de \zeta^{\nu}} + \frac{\phi^2(p_0)}{{\cal F}^2}\, \frac{\de p_0^{\mu}}{\de \zeta^{\mu}} \, \frac{\de p_0^{\nu}}{\de \zeta^{\nu}} \right], \label{eq:gmunuQ1}
\eea
where the normalization factor ${\cal F}=m {\cal A}$ denotes the maximum proper force, with ${\cal A}$ representing the maximum proper acceleration defined as ${\cal A}=c^7/G\hbar$. In this framework, ${\cal F}^2$ signifies the normalization applied to the square of the resultant proper force. During the derivation of Eq. (\ref{eq:gmunuQ1}), it was presumed that the products of mixed derivatives completely vanish.

\section{Horizontal and Vertical Lifts Adjusted Frame on Tangent Bundle $TM$}

In order to define a Sasaki metric, we split the tangent space $TM$ at a position $(x,y)$ into the horizontal and vertical subspaces associated the the base connection $\nabla$, such that
\begin{itemize}
    \item the vertical subspace, $V_{(x,y)} TM$ is identified with $T_x M$ and is spanned by the coordinate vector fields, $\dot{\partial}_{i}$, and
    \item the horizontal subspace $H_{(x,y)}TM$ is defined using the connection $\nabla$.
\end{itemize}
The horizontal lift of the base vector field is given as 
\begin{equation}
    \delta_i=\frac{\delta}{\delta_i}=\partial_i - \Gamma^k_{ij}(x) y^j \dot{\partial}_k.
\end{equation}
For horizontal and vertical subspaces, a global adjusted frame on the tangent bundle $TM$ is obtained
\begin{equation}
    [\delta_i, \dot{\partial}_i].
\end{equation}
The dual to this frame obviously reads
\begin{equation}
    [dx^i, \delta y^i],
\end{equation}
where $\delta y^i$ is expressed as
\begin{equation}
    \delta y^i=dy^i+\Gamma^i_{jk}(x)y^j dx^k.
\end{equation}
Accordingly, we find that 
\bea
    \langle dx^i, \delta y^j \rangle &=& \delta^{ij}, \\
    \langle \delta y^i, \dot{\partial}_j \rangle &=& \delta^i_{j}.
\eea
It is obvious that the horizontal basis removes the connection part of $\partial_i$ to make horizontal vectors transform like base vectors under a coordinate change.

\section{Sasaki Metric on Tangent Bundle $TM$}
\label{sec:Sasaki}

Given $(M,g)$ with components $g_{ij}$, the Sasaki metric $g^S$ on the tangent bundle $TM$ is a metric for which the horizontal and vertical subspaces are orthogonal and their inner products produce the base metric when using horizontal or vertical lifts. For the adjusted frame, It was found that 
\begin{equation}
    g^S = g_{ij}(x)dx^i \otimes dx^j+g_{ij}(x) \delta y^i \otimes \delta y^j.
\end{equation}
On the adjusted basis,  the nonvanishing components are given as
\bea
    g^S(\delta_i, \delta_j) &=& g_{ij}(x), \\
    g^S(\dot{\partial}_i, \dot{\partial}_j) &=& g_{ij}(x).
\eea
The inner product for orthogonal components reads
\begin{equation}
    g^S(\delta_i, \dot{\partial}_j)=0,
\end{equation}
where the line element is then expressed as
\begin{equation}
    (ds^2)_S=g_{ij}(x)dx^i dx^j+g_{ij}(x)\left(dy^i+\Gamma^i_{kl}y^k dx^l\right) \left(dy^i+\Gamma^j_{mn}y^m dx^n \right). \label{eq:dsSasaki}
\end{equation}
Accordingly, we conclude that this adapted form keeps the block diagonal.

\section{Coordinate Expansion and Relation to Levi--Civita Connection}

How to express the Sasaki metric $g^S$ in the coordinate frame $(\partial_i, \dot{\partial}_{i}$) rather than the adjusted frame is discussed in this section. We start with Eq. \eqref{eq:dsSasaki}. By expanding the second term, we find that
\begin{itemize}
    \item $g_{ij} dy^i dy^j$ are vertical-vertical components, 
    \item $2g_{ij}(x)\Gamma^j_{mn}y^m dy^i dx^n$ are vertical-horizontal components, and
    \item $g_{ij}(x) \Gamma^i_{mn} y^k y^m dx^l dx^n$ express the correction to the horizontal-horizontal components when seen in the coordinate basis $\partial_i$.
\end{itemize}

If one expresses the matrix of the Sasaki metric $g^S$ in the coordinate basis as $(x^i, y^j)$, then the block form reads
\begin{equation}
    \begin{bmatrix}
        A_{ij}(x,y) & B^j_i (x,y)\\
        (B^T)^i_j (x,y) & g_{ij}(x)
    \end{bmatrix},
\end{equation}
where
\bea
    A_{ij}(x,y) &=& g_{ij}(x)+g_{ab}(x)\Gamma^a_{ik}(x) \Gamma^b_{jl}(x) y^k y^l, \\
    B^j_i (x,y) &=& g_{ab}(x)\Gamma^a_{ik}(x) y^k \delta^{bj}.
\eea
Now, instead of expanding every index, we keep the adjusted-coframe form as the canonical respresentation. This is the cleanest presentation for subsequent geometric manipulations.

The Sasaki metric on cotangent bundle $T^*M$ is elaborated in next section, \ref{sec:Sasakimtr}.

\section{Sasaki Metric on Cotangent Bundle $T^*M$}
\label{sec:Sasakimtr}

The cotangent bundle $T^*M$ with the coordinates $(x^i, p_i)$ is an analogy of the Sasaki metric. In this regard, the adapted coframe can be defined as 
\begin{equation}
    \delta p_i=dp_i - \Gamma^k_{ij}(x)p_k dx^j.
\end{equation}
Also, in the adapted form, the cotangent Sasaki-type is given as
\begin{equation}
    \tilde{g}^S=g^{ij}(x)dx_i dx_j + g^{ij}(x)\delta p_i \delta p_j.
\end{equation}
For Hamiltonian or Finsler setting, the canonical symmetric quantity used on the cotangent bundle $T^*M$ is the Hessian of a Hamiltonian $H(x,p)$ with respect to $p$ so that
\begin{equation}
    G^{ij}(x,p)=\frac{\partial^2 H(x,p)}{\partial p_i \partial p_j},
\end{equation}
where the resulting metric $G^{ij}$ is placed instead of the inverse base metric.

\section{Classical Limit}

In the limit that $\alpha,\alpha',\beta \to 0$, one finds
\begin{equation}
\psi_\mu \to 0, \qquad \phi(p_0)\to 1.
\end{equation}
Also, the nonlinear connection vanishes
\begin{equation}
N^\mu_{\ \nu}\to 0.
\end{equation}
Consequently, the phase-space metric reduces to the block-diagonal form, namely
\begin{equation}
\tilde g_{AB} \to
\begin{pmatrix}
g_{\mu\nu}(x) & 0 \\
0 & g_{\mu\nu}(x)
\end{pmatrix},
\end{equation}
recovering classical general relativity.

\end{document}